\documentclass[usenatbib,useAMS english]{mnras}
\usepackage[fleqn]{amsmath}
\usepackage{array,multirow,graphicx}
\usepackage{txfonts}
\usepackage{lmodern}
\bibpunct{(}{)}{;}{a}{}{,}
\usepackage{floatrow}
\usepackage{array}
\usepackage{relsize}
\usepackage{bm}
\usepackage{listings}
\usepackage{url}
\usepackage[font=small,labelfont=bf]{caption}

\graphicspath{{./figs/}}

\newcommand{\ds}{{\sc Dark Sage}}
\newcommand{\zo}{$z\!=\!0$}
\newcommand{\HI}{H\,{\sc i}} 
\newcommand{\Htwo}{H$_{2}$}
\newcommand{\mHI}{$m_{\rm H\,{\LARGE{\textsc i}}}$} 
\newcommand{\rHI}{$r_{\rm H\,{\LARGE{\textsc i}}}$}
\newcommand{\rhalf}{$r_{\rm H\,{\LARGE{\textsc i}}}^{\rm half}$}
\newcommand{\SHI}{$\Sigma_{\rm H\,{\LARGE{\textsc i}}}$}
\newcommand{\SHIr}{$\Sigma_{\rm H\,{\LARGE{\textsc i}}}(r)$}
\newcommand{\Msol}{M$_{\odot}$} 

\newfloatcommand{capbtabbox}{table}[][\FBwidth]

\def\app#1#2{%
\mathrel{%
\setbox0=\hbox{$#1\sim$}%
\setbox2=\hbox{%
\rlap{\hbox{$#1\propto$}}%
\lower1.1\ht0\box0%
}%
\raise0.25\ht2\box2%
}%
}
\def\approxprop{\mathpalette\app\relax}

\title[Origin of the galaxy \HI~size--mass relation]{Origin of the galaxy \HI~size--mass relation}
\author[A.~R.~H.~Stevens et al.]{Adam R.~H.~Stevens,$^{1,2}$\thanks{E-mail: adam.stevens@uwa.edu.au} Benedikt Diemer,$^3$ Claudia del P.~Lagos,$^{1,2}$ Dylan Nelson,$^4$ 
\newauthor  Danail Obreschkow,$^{1,2}$ Jing Wang$^5$ and Federico Marinacci$^6$\\
$^1$International Centre for Radio Astronomy Research, The University of Western Australia, Crawley, WA 6009, Australia\\
$^2$Australian Research Council Centre of Excellence for All Sky Astrophysics in 3 Dimensions (ASTRO 3D)\\
$^3$Institute for Theory and Computation, Harvard-Smithsonian Center for Astrophysics, Cambridge, MA 02138, USA\\
$^4$Max-Planck-Institut f\"{u}r Astrophysik, D-85741 Garching, Bayern, Germany\\
$^5$Kavli Institute for Astronomy and Astrophysics, Peking University, Beijing 100871, China\\
$^6$Department of Physics \& Astronomy, University of Bologna, 40129 Bologna, Italy
}

\begin{document}

\defcitealias{gk11}{GK11}
\defcitealias{gd14}{GD14}
\defcitealias{k13}{K13}
\defcitealias{wang16}{W16}

\pagerange{\pageref{firstpage}--\pageref{lastpage}} \pubyear{2019}

\maketitle

\label{firstpage}

\begin{abstract}
We analytically derive the observed size--mass relation of galaxies' atomic hydrogen (\HI), including limits on its scatter, based on simple assumptions about the structure of \HI~discs.  We trial three generic profiles for \HI~surface density as a function of radius.  Firstly, we assert that \HI~surface densities saturate at a variable threshold, and otherwise fall off exponentially with radius or, secondly, radius squared.  Our third model assumes the \emph{total} gas surface density is exponential, with the \HI~fraction at each radius depending on local pressure.  These are tested against a compilation of 110 galaxies from the THINGS, LITTLE THINGS, LVHIS, and Bluedisk surveys, whose \HI~surface density profiles are well resolved.  All models fit the observations well and predict consistent size--mass relations.  Using an analytical argument, we explain why processes that cause gas disc truncation -- such as ram-pressure stripping -- scarcely affect the \HI~size--mass relation.  This is tested with the IllustrisTNG(100) cosmological, hydrodynamic simulation and the \ds~semi-analytic model of galaxy formation, both of which capture radially resolved disc structure.  For galaxies with $m_*\!\geq\!10^9\,{\rm M}_{\odot}$ and $m_{\rm H\,{\LARGE{\textsc i}}}\!\geq\!10^8\,{\rm M}_{\odot}$, both simulations predict \HI~size--mass relations that align with observations, show no difference between central and satellite galaxies, and show only a minor, second-order dependence on host halo mass for satellites.  Ultimately, the universally tight \HI~size--mass relation is mathematically inevitable and robust.  Only by completely disrupting the structure of \HI~discs, e.g.~through overly powerful feedback, could a simulation predict the relation poorly.
\end{abstract}



\section{Introduction}
\label{sec:intro}

The majority of our understanding surrounding the structure of cold gas in low-redshift galaxies comes from observations of the ubiquitous emission line of atomic hydrogen (\HI) at a rest-frame wavelength of $\sim\!21\,{\rm cm}$.  \HI~in galaxies is well documented to lie in rotationally supported discs that extend notably beyond optical discs from stellar emission (e.g.~\citealt{bosma81}; but see \citealt{meurer18}).  This arises because stars form in dense gaseous regions, where a more significant fraction of hydrogen is in a molecular state \citep[e.g.][]{bigiel08,leroy08}.  Meanwhile, the gas in a disc with higher specific angular momentum (farther from the global minimum of the potential well) is stable against local gravitational collapse, and so remains in an atomic state \citep{ob16,stevens18}.  Being more distant from sites of star formation, this gas can also be less prone to feedback effects \citep[although the interplay between galaxies' \HI~content and feedback is non-trivial -- see e.g.][]{crain17}.

As first highlighted by \citet{broeils97}, a key feature of \HI~discs is a \emph{genuinely} tight relation between their size and mass.  \HI~size has canonically been measured as the radius at which the surface density profile, \SHIr, drops below 1\,\Msol\,pc$^{-2}$, hereafter denoted \rHI.  
This convention arose in part because many earlier radio observations were not sensitive to \HI~column densities much lower than this (dating back to the likes of \citealt{warmels88,broeils94}).
The relation between \rHI~and integrated \HI~mass, \mHI, is a simple power law that holds over more than four decades in \mHI, with measured scatter [standard deviation in $\log_{10}\left( r_{\rm H\,{\LARGE{\textsc i}}}\right)$~from the best-fitting power law] of 0.06\,dex ($<\!15$ per cent; \citealt*{begum08,lelli16,wang16}).  The monotonic nature of this relation is often framed as meaning all galaxies have a common average \HI~surface density within \rHI~\citep[e.g.][]{broeils97,verheijen01,begum08,wang16}.  This implies there must be commonality amongst the \SHIr~profiles of all galaxy discs \citep{wang14}.

Over the last decade, it has been shown on numerous occasions that the \HI~size--mass relation is reproducible in both cosmological, hydrodynamic simulations \citep{wang14,bahe16,marinacci17,diemer19} and semi-analytic models \citep{ob09,wang14,lutz18}, although often not all of the relation's three defining values (slope, normalization, and scatter) \emph{precisely} align with the narrow empirical ranges.  The choice of prescription for how neutral hydrogen is broken into its atomic and molecular components in these models generally does not affect this outcome, even though this can change the \emph{exact} form of \SHIr~(although the results of \citealt{bahe16} appear to be an exception).  Rather, it is only in instances when implemented feedback effects are evidently too strong or interact with the interstellar-medium model in an unexpected fashion that simulated galaxies start to deviate from the \HI~size--mass relation.  For example, \citet[][see their fig.~6]{bahe16} explicitly show that galaxies containing excessively large \HI~`holes' in the EAGLE simulations steepen the predicted slope of the size--mass relation; when these galaxies are excluded, the relation returns to consistency with \citet{broeils97}.

While many works have highlighted the existence and significance of the \HI~size--mass relation, we have not yet seen a \emph{mathematically explicit} description for why the relation exists.  \citet{wang14} showed that observations, zoom-in hydrodynamic simulations, and a semi-analytic model \citep[with resolved disc structure --][]{fu13} can all produce galaxies with \SHIr~profiles of a common shape.  They comment that this commonality should explain the tightness of the \HI~size--mass relation, although it is not explicitly derived. 
In this paper, we use simple models of increasing complexity to describe galaxy discs, from which we analytically derive the \HI~size--mass relation.  Using these models, we investigate what impact disc truncation from an effect like ram pressure would have.  By weighing this against recent observational and simulated data, we discuss how the \SHIr~profiles of satellite galaxies must be altered as they are stripped.

This paper is structured as follows.  In Section \ref{sec:data}, we give a brief overview of the observations and simulations we use to support our analysis.  We then present our analytic models in Section \ref{sec:models}, deriving an \HI~size--mass relation in each case, and comparing how well these models reflect both real and simulated data.  Our models are extended in Section \ref{sec:env} to consider the effects of ram-pressure stripping.  Here, we also explore the impact of halo mass on the \HI~size--mass relation as predicted by both the TNG100 simulation and {\sc Dark Sage} semi-analytic model.  Section \ref{sec:conc} finally offers a brief conclusion.  Supplementary equations and analysis can be found in Appendices \ref{app:dists}, \ref{app:trunceq}, and \ref{sec:rhalf}.


\section{Supporting data}
\label{sec:data}

While not the main focus of this work per se, we use data from both observations and simulations to help support and/or contextualize our arguments throughout this paper.  We briefly describe them here.  Note that, where relevant, we assume $h\!=\!0.6774$, per the \citet{planck16} cosmological parameters.

\subsection{21-cm observations}
\label{ssec:obs}
There is an ever-increasing sample of galaxies in the literature that have resolved 21-cm maps, from which \HI~surface density profiles are inferred. In this paper, we use profiles from a variety of sources.  These include 16 galaxies from The \HI~Nearby Galaxy Survey \citep[THINGS;][]{walter08}, 14 from LITTLE THINGS \citep{hunter12}, 41 from The Local Volume \HI~Survey (LVHIS; data originally presented by \citealt*{ryder95,westmeier11,westmeier13}; for the complete survey, see \citealt{koribalski18}), and 39 from the Bluedisk sample \citep{wang13}.  These comprise a subset of the galaxy sample used in \citet[][hereafter W16]{wang16}.  All of these galaxies have well-resolved, inclination-corrected \SHIr~profiles, with cleanly measured \HI~sizes and masses.

The THINGS galaxies in our sample are the same subset used by \citet{og14}, which are all definitively spirals, spanning a stellar-mass range of $2.5 \times 10^9$ -- $1.6 \times 10^{11}\,{\rm M}_{\odot}$.
The galaxies we use from LITTLE THINGS are the same subset as in \citet*{butler17}, covering $1.4 \times 10^6$ -- $2.0 \times 10^8\,{\rm M}_{\odot}$ in stellar mass, and are morphologically classified as dwarf irregulars.
The LITTLE THINGS and Bluedisk galaxies are predominantly isolated, while the LVHIS galaxies mainly occupy a subgroup near the Sculptor Group. 
The LVHIS galaxies we include are selected to have \rHI~greater than 1.5 times the major axis of the PSF\footnote{Point Spread Function} ellipse, ensuring the disc profiles are sufficiently resolved (the other data more than meet this criterion already).
Most galaxies from LVHIS and Bluedisk are classified as spirals.
All galaxies in our sample are at $z\!\simeq\!0$.
We refer the reader to the specific papers where the data are presented for further details.
While we cannot guarantee that this sample is representative of \emph{all} galaxies in the local Universe (in fact, it is biased towards rotation-dominated systems), we take and analyse the data as they are.  Our simulated data help compensate by offering volume-limited samples that are orders of magnitude larger in galaxy number.

For the Bluedisk galaxies, we calculate \mHI~by numerically integrating the full surface density profile of each galaxy.  These \HI~masses are $\sim$15 per cent larger than the `true' \mHI~values given in \citet{wang13}, which were only integrated out to a finite surface density.  Any pre-measured \mHI~quantities for the other galaxies are consistent with numerically integrating their profiles.

\subsection{IllustrisTNG}
\label{ssec:tng}

IllustrisTNG\footnote{Illustris: The Next Generation} comprises a suite of cosmological, magnetohydrodynamic simulations of various volumes and resolutions, run with the {\sc arepo} code \citep{springel10}.  In this paper, we use the main TNG100 simulation\footnote{TNG100 (and TNG300) have recently been made publicly available \citep{nelson19}.} \citep{pillepich18b,nelson18,marinacci18,naiman18,springel18}, with a periodic box of length $75\,h^{-1} \! \simeq \! 110\,{\rm cMpc}$, containing $1820^3$ dark-matter particles of mass $7.5 \! \times \! 10^6\,{\rm M}_{\odot}$, and $1820^3$ initial baryonic elements of typical mass $1.4 \! \times \! 10^6\,{\rm M}_{\odot}$.  TNG simulations include subgrid models to follow gas cooling, star formation, growth of massive black holes, and feedback from both stars and active galactic nuclei \citep{weinberger17,pillepich18a}.  
Black-hole feedback removes gas from its immediate neighbourhood ($\lesssim\!1$\,kpc from the centre), while supernova feedback removes gas everywhere according to the  local star formation rate (on $\sim$500\,pc scales) and induced mass-loading factor.  
The simulations and methods are based on the earlier Illustris project \citep{vogelsberger13,vogelsberger14a,vogelsberger14b,genel14,torrey14}.

Gas cells in the simulation are post-processed to calculate their mass fractions in the form of atomic and molecular hydrogen \citep{diemer18,stevens19}.  We present results from three methods, based on the works by \citet{gk11}, \citet{k13}, and \citet{gd14}.  We refer the reader to \citet{stevens19} and references therein for full details on the methodology; all properties in this paper follow the `inherent' method, meaning only particles/cells associated with the {\sc subfind} object \citep{springel01,dolag09} that also meet the spherical-aperture criterion of \citet{stevens14} are included.  
\HI~radii are derived by building one-dimensional \HI~surface density profiles, using cylindrical annuli with an axis parallel to the galaxy's angular-momentum vector (computed exclusively from stellar particles), and linearly interpolating the exact position where these profiles drop below $1\,{\rm M}_{\odot}\,{\rm pc}^{-2}$.

For this work, we include galaxies at \zo~with stellar masses above $10^9\,{\rm M}_{\odot}$, \HI~masses above $10^8\,{\rm M}_{\odot}$, and \HI~radii greater than the minimum gas softening length of 190\,pc (both these \HI~requirements only needed to be satisfied for one of the three \HI/\Htwo~prescriptions).   Our resulting TNG100 sample totals $\sim$15\,000 galaxies; the sample size would be $\sim$20\,000 with just the stellar-mass cut alone.
The added \mHI~and \rHI~cuts somewhat bias us towards star-forming galaxies; these cuts reduce the total passive fraction from $\sim$28 to $\sim$5 per cent, where we define a `passive' galaxy as one with a specific star formation rate $<\!10^{-11}\,{\rm yr}^{-1}$ (based on the gas cells' instantaneous star formation rates).  In practice, a passive TNG100 galaxy often has a star formation rate of zero.


\subsection{D{\small ARK} S{\small AGE}}
\label{ssec:ds}

\ds~is a semi-analytic model of galaxy formation originally developed by \citet*{stevens16}.  Its stand-out features include a comprehensive consideration of the angular momentum of galaxy discs.  Each disc is broken into a series of 30 annuli \citep[similar to][]{fu10} whose edges are fixed in their specific angular momentum \citep[\`{a} la][]{stringer07} and spaced logarithmically.  The net orientation and magnitude of gas and stellar discs' specific angular momenta are tracked and continuously updated based on the astrophysical processes considered.  Among others, these processes include gas cooling, star formation and stellar feedback, and the growth and feedback of black holes, where each of these are calculated on an annulus-by-annulus basis.
For example, stellar feedback only reheats gas out of the same annulus where the precursory star formation took place, while quasar winds initially remove gas from the central annulus and can extend to outer annuli based on the energy involved.
The publicly available \ds~code\footnote{\url{https://github.com/arhstevens/DarkSage}} (and many of the physical prescriptions) is based on \citet{croton06,croton16}.  For a more thorough overview of semi-analytic models in general, see e.g.~\citet{baugh06,somerville15}.

\ds~accounts for the effects of ram-pressure stripping on satellite galaxies at a level of detail beyond most other semi-analytic models.  Provided a sufficient amount of hot gas around a satellite is lost, a prescription based directly on \citet{gunn72} is applied to each annulus individually.  Where ram pressure exceeds the local restoring force per unit area, all gas in the satellite's annulus is transferred to the intra-halo medium (i.e.~the hot component associated the corresponding central galaxy).  Barring extreme circumstances, the local restoring force of discs decreases with radius.  As such, ram-pressure stripping in \ds~leads to the continual truncation of gas discs.  Satellites are also denied cosmological accretion of gas, and have their hot-gas reservoir gradually depleted through tidal or ram-pressure stripping (manifesting as starvation/strangulation -- cf.~\citealt*{larson80}).  Satellite galaxy discs can still accrete \emph{from} that hot gas though, where the specific-angular-momentum vector of that gas is fixed at infall.

We use the \citet{stevens18} version of \ds~in this work.  This was run on the Millennium simulation \citep{springel05}.  Even though the cosmology assumed in this simulation \citep{spergel03} differs from \emph{Planck}, to be consistent with our other results, we use $h\!=\!0.6774$ for our \ds~results.  We otherwise maintain the galaxy properties as they are in \citet{stevens18}, meaning there is no rescaling to account for the other cosmological parameters \citep[but see][]{angulo10}.  The prescription for the \HI/\Htwo~breakdown used in this version of the model is based on \citet{mckee10}.  Taking the centre of each annulus as its position for the galaxies' \SHIr~profiles, we linearly interpolate between the outermost annulus with $\Sigma_{\rm H\,{\LARGE{\textsc i}}} \! > \! 1\,{\rm M}_{\odot}\,{\rm pc}^{-2}$ and the next to obtain \rHI.  Because each consecutive annulus edge has 40 per cent higher specific angular momentum, the separation between the annuli where \rHI~is measured is typically $\sim \! 0.4\,r_{\rm H\,{\LARGE{\textsc i}}}$.  We only analyse redshift-zero \ds~galaxies in this paper that occupy (sub)haloes that have been composed of at least 100 particles (equivalent to a halo mass of $8.6  \times 10^{10}\,h^{-1}\,{\rm M}_{\odot}$) at some point in their merger-tree history, and whose stellar masses are above $10^9\,{\rm M}_{\odot}$ and \HI~masses above $10^8\,{\rm M}_{\odot}$ at \zo.  This leaves us with 4.3 million \ds~galaxies.


\section{Disc models and derivations of the \HI~size--mass relation}
\label{sec:models}

\begin{figure}
\centering
\includegraphics[width=\textwidth]{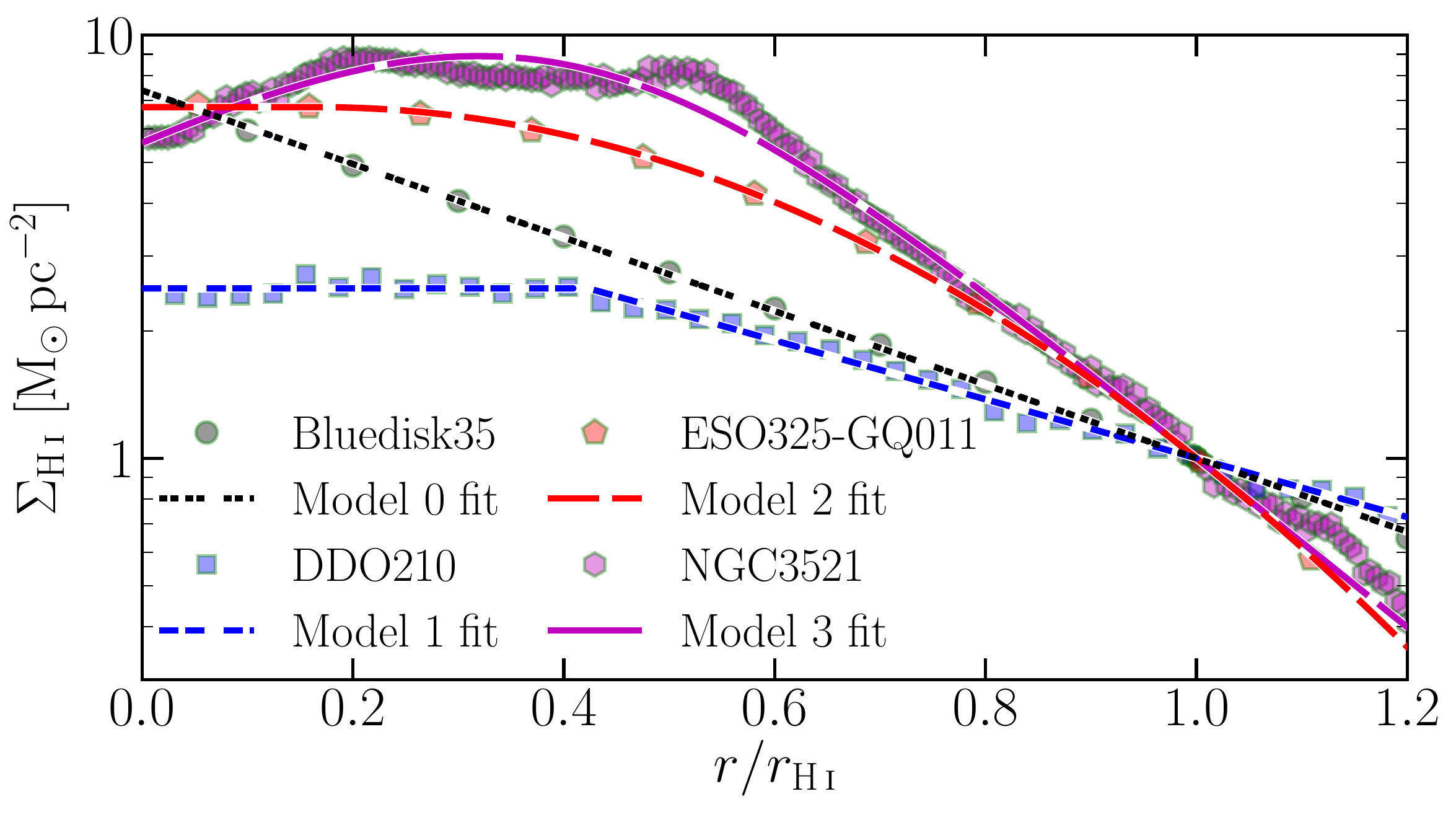}
\caption{\HI~surface density profiles of four galaxies from our observational sample (points).  These have been hand-picked to show examples of when each of our analytic models in Section \ref{sec:models} is an accurate representation of reality; each line is the best fit of a different model to a different galaxy, with colour indicating which line is a fit to which data.}
\label{fig:profeg}
\end{figure}

In this section, we explore several models of progressively increasing complexity for the one-dimensional distribution of \HI~in galaxy discs.  For each model, we show an example \SHIr~profile in Fig.~\ref{fig:profeg}, which is accompanied by a real example galaxy whose observed \HI~surface density profile is well described by that model.  We will show that regardless of how much detail is added to the disc profiles, one can always mathematically derive a tight \HI~size--mass relation that matches observations.  Note that, throughout parts of this section, we use a bar to denote when surface densities and radii have been normalized:
\begin{subequations}
\label{eq:bar}
\begin{equation}
\bar{r}_x \equiv r_x / r_{\rm H\,{\LARGE{\textsc i}}}\,,
\end{equation}
\begin{equation}
\bar{\Sigma}_x \equiv \Sigma_x / \left(1\,{\rm M}_{\odot}\,{\rm pc}^{-2}\right)\,,
\end{equation}
\end{subequations}
where $x$ represents any subscript.

Many parameters and expressions are introduced in this section.  As a point of reference, we summarize the definitions and typical values of the key parameters of all our models in Table \ref{tab:params}.

\begin{table*}
\centering
\begin{tabular}{l l c c c c c} \hline\hline
Parameter & Definition & Model & Mathematically  & A priori  & Full range of & 68\% interval \\
 & & & allowed values &  expectation &  fits to obs. & of obs.  \\\hline\hline
 & Maximum/saturation  & 0 &   &  & $(2.5,35.2)$ & $(7.0,22.3)$ \\
$\bar{\Sigma}_0$ &  \HI~surface density, & 1 &  $>\!1.0$ & $\sim$2--10 & $(1.3,22.4)$ & $(3.2,8.5)$ \\
 & normalized by $1\,{\rm M}_{\odot}\,{\rm pc}^{-2}$ & 2 &  &  & $(1.5,14.8)$ & $(3.2,8.7)$  \\\hline
\multirow{2}{*}{$\bar{r}_b$} & Saturation break radius, & 1 & \multirow{2}{*}{[0,\,1]} & \multirow{2}{*}{$\sim$0--0.8} & $(0.01,0.83)$ & $(0.25,0.65)$ \\
 & normalized by \rHI & 2 & & & $[0,0.65)$ & $(0,0.46)$ \\\hline
\multirow{2}{*}{$\bar{\Sigma}_{\rm 0,H}$} & Normalized maximum & \multirow{2}{*}{3} & \multirow{2}{*}{$\geq\!4.22$} & \multirow{2}{*}{$\sim$10--1000} & \multirow{2}{*}{$[4.22,432.7)$} & \multirow{2}{*}{$(18.5,161.6)$}  \\
 & \HI+\Htwo~surface density & & & & \\\hline
\multirow{2}{*}{$\bar{r}_d$} & Normalized exponential & \multirow{2}{*}{3} & \multirow{2}{*}{$>\!0$} & \multirow{2}{*}{$\sim$0.1--1} & \multirow{2}{*}{$(0.16,0.72)$} & \multirow{2}{*}{$(0.19,0.35)$} \\
 & scale radius for \HI+\Htwo & & & & \\\hline\hline
\end{tabular}
\caption{Summary of the parameters defining our analytic disc models, described in Sections \ref{ssec:model0}--\ref{ssec:model3}.  The `mathematically allowed values' for models 0, 1, and 2 come directly from the parameters' definitions.  For model 3, these limits are derived under the requirement that \SHIr~is always finite and real; $\bar{r}_d$ actually has stricter upper and lower limits that depend on $\bar{\Sigma}_{\rm 0,H}$ (see Equations \ref{eq:ll} \& \ref{eq:ul}).  The \emph{a priori} expectations are loosely based on previous works (\citealt{bigiel08,leroy08,stevens16}; \citetalias{wang16}). We quote both the full and 16th--84th percentile ranges of the best-fitting values to our sample of observed $\bar{\Sigma}_{\rm H\,{\LARGE{\textsc i}}}(\bar{r})\,\bar{r}$ profiles (see Section \ref{ssec:profiles}).}
\label{tab:params}
\end{table*}


\subsection{Model 0: pure exponential}
\label{ssec:model0}
Let us begin with the simple, canonical assumption that all discs are exponential.  To first order, except perhaps towards the very centre of galaxies \citep[e.g.][]{stevens17}, both stellar and gaseous galaxy discs are observationally known to follow exponential profiles for many galaxies \citep{dev59,bigiel12}, for which a theoretical explanation has been discussed in several works \citep[e.g.][]{freeman70,dutton09,elmegreen13}.  Let us further assert that \HI~discs \emph{specifically} are also exponential.  While this assertion is not generically supported by observations (and is therefore incomplete), it will serve as a starting point in our exploration of \HI~disc models, and hence is why we refer to this as `model 0' (effectively, we are ignoring the existence of molecular gas).  With this,
\begin{equation}
\Sigma_{\rm H\,{\LARGE{\textsc i}}}(r) = \Sigma_0 \exp\left(-r/r_s \right)\,,
\label{eq:SHIr_0}
\end{equation}
where $r_s$ is the exponential scale radius and $\Sigma_0$ is the central \HI~surface density.  The total \HI~mass is then
\begin{subequations}
\begin{equation}
\label{eq:integral}
m_{\rm H\,{\LARGE{\textsc i}}} \equiv {2\pi} \int_0^{\infty} \Sigma_{\rm H\,{\LARGE{\textsc i}}}(r)\, r\, {\rm d}r
\end{equation}
\begin{equation}
\Rightarrow m_{\rm H\,{\LARGE{\textsc i}}} = 2\pi\, \Sigma_0\, r_s^2\,.
\end{equation}
\end{subequations}
In reality, an \HI~disc would not extend to infinity; at some point, one would reach the ionized intergalactic medium or another object.  Because the integral is convergent though, we assume (throughout this paper) that \HI~discs extend to sufficiently large radii such that integrating to infinity is a valid approximation.\footnote{For $\bar{\Sigma}_0 \! = \! 5$, this approximation is accurate to 10 per cent if the disc actually only extends to $\sim$2.4\,\rHI, and is accurate to 1 per cent if it extends to $\sim$4.1\,\rHI.  Higher values of $\bar{\Sigma}_0$ converge at lower radii (and vice versa).}

We should also recognize that $r_s$ can be rewritten in terms of \rHI.  That is, for an exponential profile, it must be true that
\begin{equation}
r_{\rm H\,{\LARGE{\textsc i}}} = \ln\!\left({\frac{\Sigma_0}{\Sigma_c}}\right) r_s\,,
\label{eq:rHI_0}
\end{equation}
where $\Sigma_c \! = \! 1\,{\rm M}_{\odot}\,{\rm pc}^{-2}$, as per the definition of \rHI~(although, in principle, one could define \rHI~at a different threshold $\Sigma_c$, e.g.~as explored in fig.~4 of \citetalias{wang16}).  After some short algebra, one can simply solve for \rHI~in terms of \mHI:
\begin{subequations}
\begin{equation}
r_{\rm H\,{\LARGE{\textsc i}}} = f(\Sigma_0)\, m_{\rm H\,{\LARGE{\textsc i}}}^{0.5}~,
\end{equation}
\begin{equation}
f(\Sigma_0) = \left(2\pi\, \Sigma_0\right)^{-0.5} \ln\left(\frac{\Sigma_0}{\Sigma_c}\right)\,.
\end{equation}
\end{subequations}

With the above, we have already derived an \HI~size--mass relation with a normalization (in log--log space) that depends solely on $\Sigma_0$.  Observations have shown that it is rare for \SHI~to exceed $9\,{\rm M}_{\odot}\,{\rm pc}^{-2}$ in local galaxies \citep[on scales of $\sim$750\,pc --][]{bigiel08}; at higher surface densities, hydrogen tends to be sufficiently cool and self-shielded to promote the formation of molecules and prevent their photodissociation.  But theoretically, the physical limit on \SHI~for a given galaxy depends on metallicity \citep[e.g.][]{schaye01,kmt09}, so higher values of $\Sigma_0$ should be possible.  For now, we take $10\,{\rm M}_{\odot}\,{\rm pc}^{-2}$ as the fiducial value for $\Sigma_0$ in our model.  Plugging this in gives $f\!\left(10\,{\rm M}_{\odot}\,{\rm pc}^{-2}\right) \! = \! 0.29\,{\rm pc}\,{\rm M}_{\odot}^{-0.5}$, or equivalently
\begin{equation}
\log_{10}\!\left( \frac{D_{\rm H\,{\LARGE{\textsc i}}}}{{\rm kpc}}\right) = 0.5\,\log_{10}\!\left(\frac{m_{\rm H\,{\LARGE{\textsc i}}}}{{\rm M}_{\odot}}\right) - 3.236
\label{eq:w2e}
\end{equation}
($D_{\rm H\,{\LARGE{\textsc i}}} \! \equiv \! 2\,r_{\rm H\,{\LARGE{\textsc i}}}$).  This expression is directly comparable to equation 2 of \citetalias{wang16} and highlights the closeness in both the slope (a best fit from \citetalias{wang16} of $0.506\!\pm\!0.003$) and intercept ($-3.293 \! \pm \! 0.009$) that is empirically derived from observations.

\begin{figure}
\centering
\includegraphics[width=\textwidth]{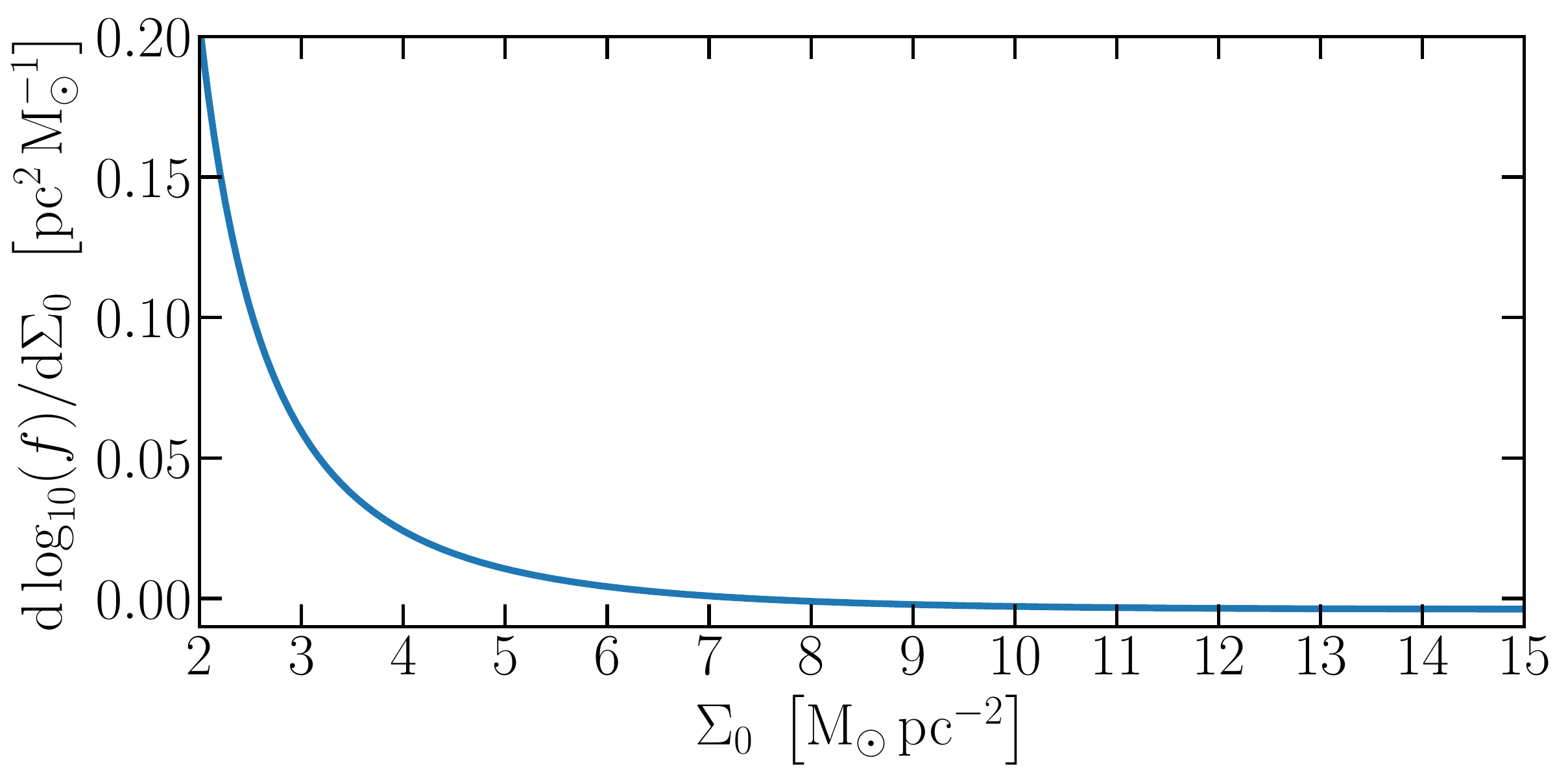}
\caption{Equation (\ref{eq:deriv}) -- the sensitivity of model 0 to its solitary parameter.  For a population of galaxies, the mean value of $f$ would represent the normalization of the \HI~size--mass relation (with slope 0.5).  The fact that the derivative of $f$ only weakly depends on $\Sigma_0$ for the majority of allowable $\Sigma_0$ values implies that the scatter in the \HI~size--mass relation cannot be large.}
\label{fig:deriv}
\end{figure}

The final characteristic trait of the \HI~size--mass relation is its small scatter (0.06\,dex).  For model 0, any scatter must come from variation in $\Sigma_0$.
Typically, the \HI~surface densities of late-type galaxies reach a maximum value anywhere from $\sim$3 to $\sim$9 \Msol\,pc$^{-2}$, while the maxima for some early-type galaxies have been observed to be even lower \citepalias[see fig.~2 of][]{wang16}.  To explicitly show that variations in $\Sigma_0$ may only lead to a small scatter, we need to differentiate (the logarithm of) $f(\Sigma_0)$.  It is straightforward to find
\begin{equation}
\frac{{\rm d} \log_{10}(f)}{{\rm d} \Sigma_0} = \frac{\log_{10}({\rm e})}{\Sigma_0} \left[\frac{1}{\ln\left(\Sigma_0/\Sigma_c\right)} - \frac{1}{2} \right]\,.
\label{eq:deriv}
\end{equation}
Fig.~\ref{fig:deriv} visualizes this derivative.  
The fact that this derivative is $\ll\!1$ for all realistic values of $\Sigma_0$, means that $f$ only depends weakly on $\Sigma_0$.
If, for example, the probability distribution function of $\Sigma_0$ for galaxies were a uniform distribution extending from 2 to $10\,{\rm M}_{\odot}\,{\rm pc}^{-2}$, then the predicted scatter in the \HI~size--mass relation would be 0.037\,dex.  Extending the upper end of this range or applying a probability distribution function that peaks at mid values of $\Sigma_0$ would only decrease the value of this prediction.

Because model 0 is incomplete, our next three models are the ones we give proper attention to throughout the rest of this paper.  Naturally though, the addition of a second parameter to the models means it is not as straightforward to explicitly derive the tightness of the \HI~size--mass relation as it was under model 0. 


\subsection{Model 1: saturated exponential}
\label{ssec:model1}

Let us now include a simple consideration of the presence of molecular gas in the disc.
We no longer assume that \HI~follows an exponential surface density profile, but instead that \emph{all} neutral gas in a disc does \citep[which is roughly consistent with a large variety of observed \HI+\Htwo~profiles analysed by][]{bigiel12}.  We then assume that below a threshold gas surface density, $\Sigma_0$, all hydrogen is in the form of \HI.  For gas at higher density, the contribution from \HI~saturates at $\Sigma_0$, where the remaining hydrogen is molecular (\Htwo).  By defining the `break radius', $r_b$, as the radius at which \HI~saturation extends to, we can formally write the \HI~surface density of our model disc as
\begin{equation}
\Sigma_{\rm H\,{\LARGE{\textsc i}}}(r) = 
\left\{
\begin{array}{l r}
\Sigma_0, & r \leq r_b\\
\Sigma_0 \exp\left[-r_s^{-1} (r-r_b) \right], & r > r_b
\end{array}
\right.\,,
\label{eq:SHIr}
\end{equation}
The relationship between $r_s$ and \rHI~must be updated from model 0, where now
\begin{equation}
r_{\rm H\,{\LARGE{\textsc i}}} = r_b + \ln\left({\frac{\Sigma_0}{\Sigma_c}}\right) r_s\,.
\label{eq:rHI}
\end{equation}
In introducing the normalizing bar (Equation \ref{eq:bar}), we can then rearrange Equation (\ref{eq:rHI}) to obtain
\begin{equation}
\bar{r}_s \equiv \frac{r_s}{r_{\rm H\,{\LARGE{\textsc i}}}} = \frac{1-\bar{r}_b}{\ln\left(\Sigma_0/\Sigma_c \right)}\,.
\label{eq:xi}
\end{equation}
The model is hence dependent on two parameters: $\Sigma_0$ and $\bar{r}_b$.  Note that setting $\bar{r}_b$ to 0 reduces this back to model 0.  As such, model 1 should always give an equally good or better fit to observed or simulated data than model 0.

Substituting Equations (\ref{eq:SHIr} \& \ref{eq:xi}) into the integral of Equation (\ref{eq:integral}) and solving for \rHI, we derive the size--mass relation for model 1:
\begin{equation}
r_{\rm H\,{\LARGE{\textsc i}}} = \sqrt{\frac{m_{\rm H\,{\LARGE{\textsc i}}}}{\pi\,\Sigma_0\,\left[\bar{r}_b^2 + 2\,\bar{r}_s\,(\bar{r}_s+\bar{r}_b) \right]}}\,.
\label{eq:sizemass}
\end{equation}
Assuming neither $\bar{r}_b$ nor $\bar{r}_s$ carry an implicit dependence on \mHI~(corresponding to self-similar surface density profiles), our simple model maintains a predicted slope of 0.5 for this fit.  The terms in the denominator of Equation (\ref{eq:sizemass}) set the normalization.  We can then try to associate the (small) scatter in the relation to variations in $\bar{r}_b$ and $\Sigma_0$.

\begin{figure}
\centering
\includegraphics[width=\textwidth]{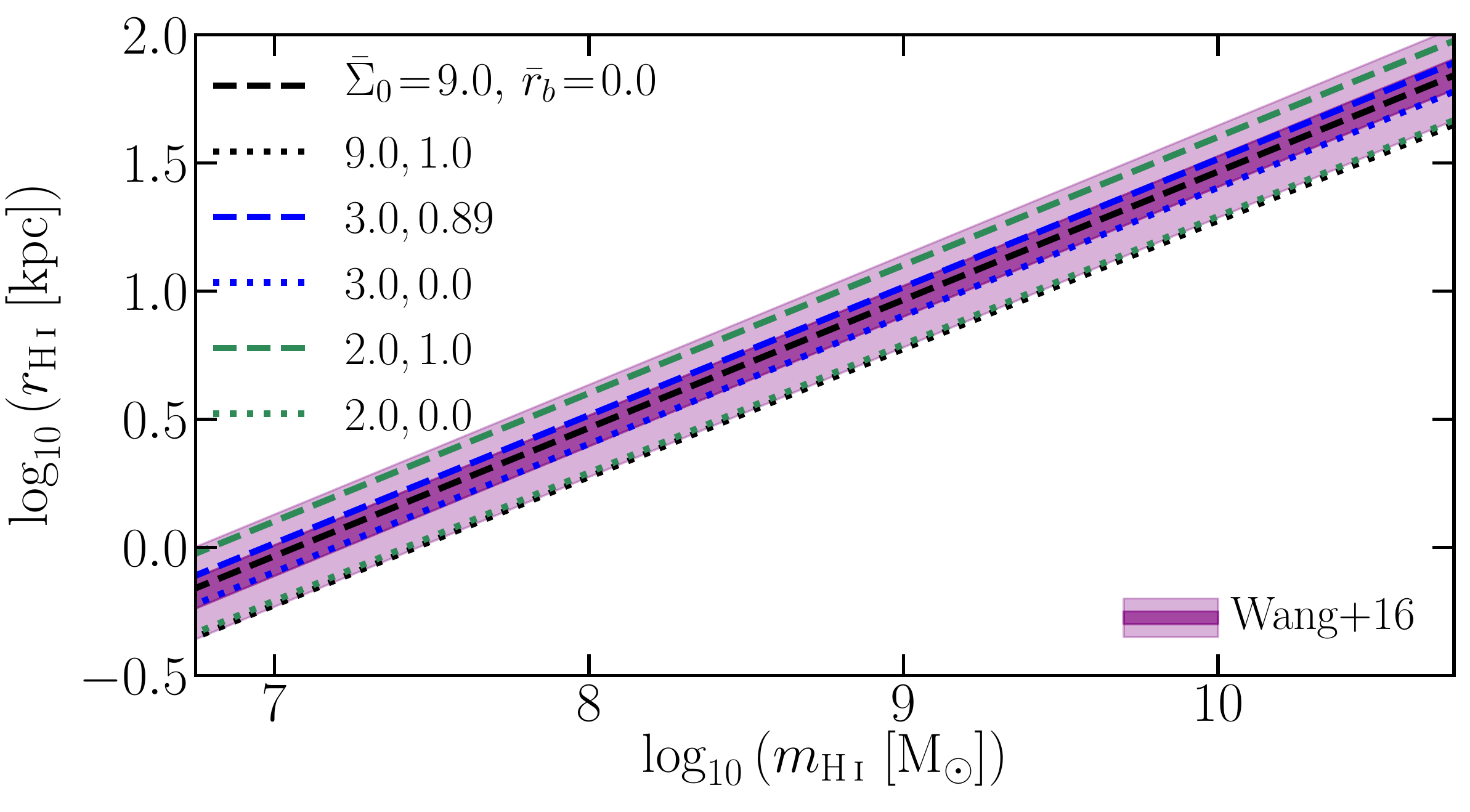}
\includegraphics[width=\textwidth]{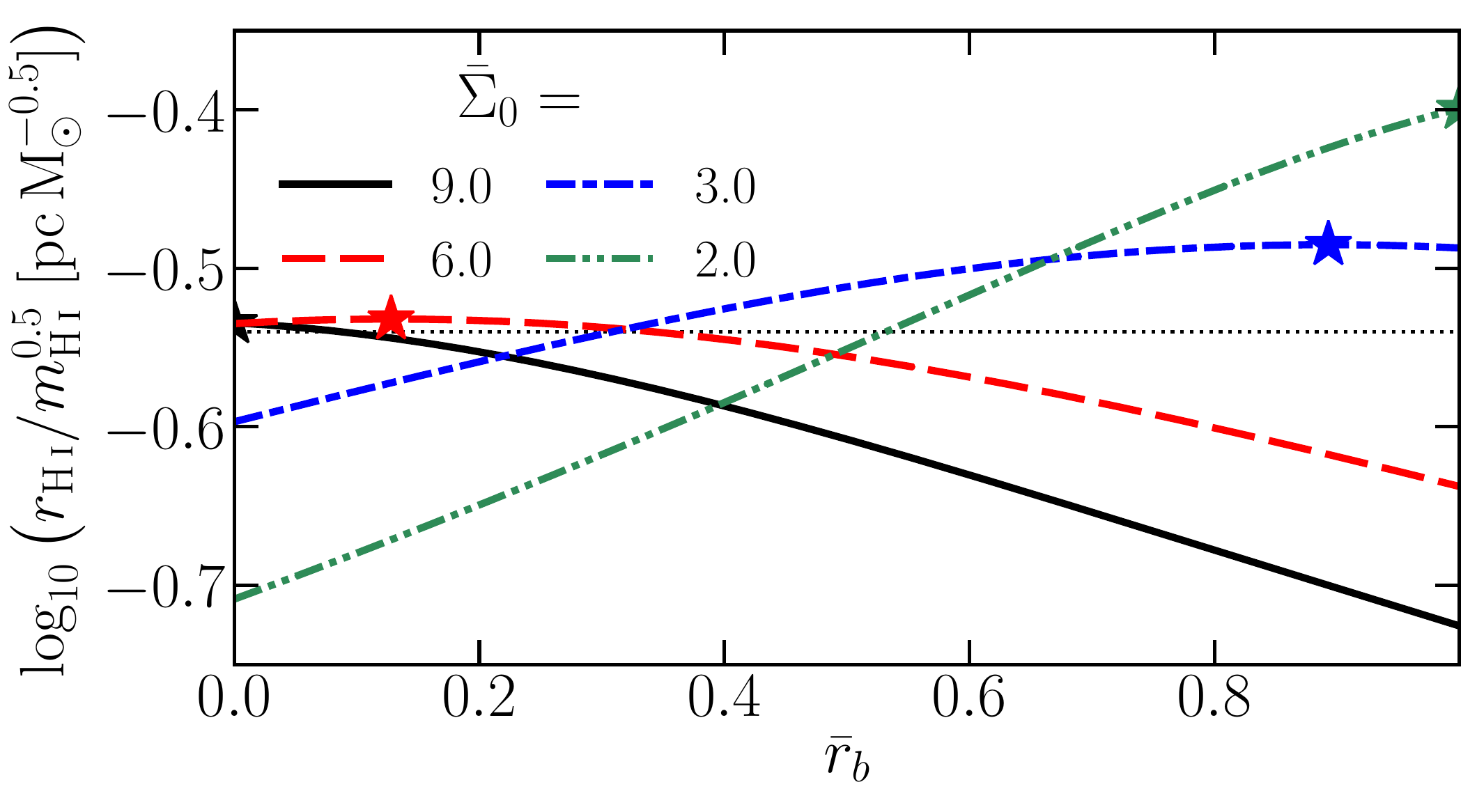}
\caption{Top panel: variation in the \HI~size--mass relation of model 1 for example parameter sets  (Equation \ref{eq:sizemass}).  We highlight several cases of $\Sigma_0$, showing the full vertical range covered $\forall \bar{r}_b \! \in \! [0,1]$ at that $\Sigma_0$.  Compared is the best-fitting relation from observational data \citepalias{wang16}; the deeper shaded region shows the $1\sigma$ scatter around the relation, and the lighter region is the $3\sigma$ scatter.  Bottom panel: a more detailed depiction of how much a model-1 line in the top panel would vertically move if $\bar{r}_b$ were varied for several examples of fixed $\Sigma_0$.  Starred points indicate where the curves reach their maximum.  The thin, dotted, horizontal line signifies zero displacement from the observed relation.}
\label{fig:model}
\end{figure}

Certainly, we must uphold $\Sigma_0 \! > \! \Sigma_c$.  And by definition in our model, $\bar{r}_b$ is restricted to the range $[0,1]$.
With these restrictions in mind, we show the allowable scatter in the size--mass relation of our model in the top panel of Fig.~\ref{fig:model}.  We cover three values of $\Sigma_0$ that have different sensitivities to $\bar{r}_b$ for relating size and mass.  For each $\Sigma_0$, we display the full range of variation in \HI~size at fixed mass for all values of $\bar{r}_b$.  In two cases, this simply means taking the extremes of $\bar{r}_b \! = \! 1$ and 0, i.e.~where \SHIr~is a top-hat%
\footnote{When $\bar{r}_b\!=\!1$, $\Sigma_{\rm H\,{\LARGE{\textsc i}}}(r) \! = \! \Sigma_0$ until a radius where it drops to zero.  This radius is also \rHI, as it is the largest radius where $\Sigma_{\rm H\,{\LARGE{\textsc i}}}(r) \! > \! \Sigma_c$.}
and pure exponential, respectively.  For $\Sigma_0 \! = \! 9\,{\rm M}_{\odot}\,{\rm pc}^{-2}$, \rHI~is smallest for $\bar{r}_b \! = \! 1$.  For $\Sigma_0 \! = \! 2\,{\rm M}_{\odot}\,{\rm pc}^{-2}$, it is the opposite: \rHI~is smallest for $\bar{r}_b \! = \! 0$.  This is because \mHI~is found by integrating \SHIr\,$r$ out to $\infty$.  Lower $\bar{r}_b$ \emph{and} lower $\Sigma_0$ each lead to a shallower \SHIr~profile beyond \rHI, meaning the mass contribution beyond \rHI~is greater.  For in-between values of $\Sigma_0$, the maximum radius at fixed mass is found at intermediate values of $\bar{r}_b$ (e.g. at $\bar{r}_b\!\simeq\!0.89$ for $\bar{\Sigma}_0\!=\!3.0$).  We highlight this in the bottom panel of Fig.~\ref{fig:model}, which is another way of showing Equation (\ref{eq:sizemass}).

At this point, one could already argue that the slope, normalization, and scatter of the \HI~size--mass relation are all mathematically inevitable.  To properly claim this though, we need to more closely analyse how representative Equation (\ref{eq:SHIr}) is of observed \SHIr~profiles. As we show and discuss in Section \ref{ssec:profiles}, model 1 is not \emph{always} a sufficient representation of reality.  Moving forward, it is therefore helpful to have further models to compare, which we present next.  We also cannot give a numerical prediction for the slope and scatter of the \HI~size--mass relation from any model without knowing how its parameter space should be occupied.  This can be inferred from the best-fitting parameter values to observations though: a task left for Section \ref{ssec:infer}.


\subsection{Model 2: empirical}
\label{ssec:model2}

We have found that many of the observed \HI~profiles in our galaxy sample follow a common shape that is more akin to falling off exponentially with radius squared (i.e.~a Gaussian), rather than just radius.  \citet{martinsson13b} also note that a Gaussian describes \SHIr~well for a completely different sample of observed galaxies.  For model 2, we therefore assert that this can be described analytically as
\begin{equation}
\Sigma_{\rm H\,{\LARGE{\textsc i}}}(r) = 
\left\{
\begin{array}{l r}
\Sigma_0, & r \leq r_b\\
\Sigma_0 \exp\left[-r_S^{-2} (r-r_b)^2 \right], & r > r_b
\end{array}
\right.\,,
\label{eq:model2}
\end{equation}
where we have maintained the option for the profile to be saturated out to $r_b$ from model 1. Following the same procedure in Section \ref{ssec:model1}, we can derive the size--mass relation for this as
\begin{subequations}
\label{eq:sizemass2}
\begin{equation}
r_{\rm H\,{\LARGE{\textsc i}}} = \sqrt{\frac{m_{\rm H\,{\LARGE{\textsc i}}}}{\pi\,\Sigma_0\,\left[\bar{r}_b^2 + \bar{r}_S\,(\bar{r}_S+\sqrt{\pi}\bar{r}_b) \right]}}\,,
\end{equation}
\begin{equation}
\bar{r}_S = \frac{1-\bar{r}_b}{\sqrt{\ln(\Sigma_0 / \Sigma_c)}}\,.
\label{eq:xip}
\end{equation}
\end{subequations}
The two parameters defining model 2 are the same as model 1 ($\Sigma_0$ and $\bar{r}_b$ -- they are just folded into different overall profiles).  The parameter space is therefore restricted in the same way.  Again, using observations to inform how this parameter space should be distributed, we infer a predicted slope and scatter for the model-2 \HI~size--mass relation in Section \ref{ssec:infer}.

We acknowledge that we have not offered a physical justification for Equation (\ref{eq:model2}). We have simply found it to empirically fit the observed \HI~profiles better than either model 1 or model 3 (introduced below) in 42 per cent of cases.  For 32 per cent of the observed profiles, the best-fitting model-2 $\bar{r}_b$ is 0. One therefore need not invoke \HI~saturation for those cases, meaning these would be well described by a one-parameter profile (akin to a variant of model 0).  We present and discuss profile fits to observations further in Section \ref{ssec:profiles}.

The danger of a Gaussian-like \SHIr~profile is that it is possible for this shape to be artificially induced by beam-smearing; the observed profile of a galaxy is a convolution of its true profile with the beam response, where the latter is well described by a Gaussian.  Many of the observed galaxies that are best represented by model 2 are the less well-resolved galaxies from LVHIS and Bluedisk.  While this should certainly be kept in mind when interpreting the general applicability of model 2, we remind the reader that the \emph{most} poorly resolved galaxies were not included in our analysis, and we note that there are galaxies from all contributing surveys to our sample that are best described by model 2.  As we will show in Section \ref{ssec:sims}, model 2 also fits many profiles from simulations well, which have not had beam-smearing effects added.


\subsection{Model 3: theoretical pressure law}
\label{ssec:model3}

For our final model, we maintain the assumption that cold-gas discs are broadly described by an exponential profile.  We then follow the idea of \citet{blitz04,blitz06} that the fraction of hydrogen at a given radius in the form of \HI~depends on the mid-plane pressure of the disc.  Using this idea, \citet[][see their equations 10 \& 11]{ob09} explicitly derive a generic \HI~profile for galaxies that still depends on an exponential scale length and the total gas and stellar mass of the disc ($m_{\rm gas}$ and $m_{\rm *,disc}$, respectively):
\begin{subequations}
\label{eq:model3}
\begin{equation}
\Sigma_{\rm H\,{\LARGE{\textsc i}}}(r) = \frac{\Sigma_{\rm 0,H}\, \exp\!\left(-r / r_d \right)}{1 + R_0\, \exp\!\left(-1.6\,r / r_d \right)}\,,
\end{equation}
\begin{equation}
R_0 = \left[K\, r_d^{-4}\, m_{\rm gas} \left(m_{\rm gas} + \langle f_{\sigma} \rangle m_{\rm *,disc} \right) \right]^{0.8}\,,
\end{equation}
\end{subequations}
where $K\!\equiv\!11.3\,{\rm m}^4\,{\rm kg}^{-2} \! = \! 4.39\!\times\!10^{-5}\,{\rm pc}^4\,{\rm M}_{\odot}^{-2}$, $\langle f_{\sigma} \rangle$ is the mean vertical velocity dispersion ratio of gas to stars in the disc, and $\Sigma_{\rm 0,H} \! \equiv \! \Sigma_{\rm H\,{\LARGE{\textsc i}}}(0) + \Sigma_{\rm H_2}(0)$.
Many assumptions go into this expression, including an empirical scaling for the pressure law \citep{leroy08}, that pressure follows the radial function of \citet{elmegreen89}, that gas velocity dispersion is a constant, that stellar discs have exponential surface density profiles with a scale length equal to $r_d/2$, and that stellar velocity dispersion decays exponentially with $r_d$.

It is useful to recognize that $m_{\rm gas}$ is not an independent parameter in Equation (\ref{eq:model3}), as it is directly connected to $\Sigma_{\rm 0,H}$:
$m_{\rm gas} = 2\pi\, r_d^2\, X^{-1}\, \Sigma_{\rm 0,H}$
(where $X \! \simeq \! 0.76$ is the mass fraction of gas that is hydrogen).
By simply defining a new quantity that also encapsulates the constants and remaining variables in Equation (\ref{eq:model3}),
\begin{equation}
\kappa \equiv \left[4.39\!\times\!10^{-5}\,\left(\frac{2\pi}{X}\right)^2\,\left(1 + \frac{\langle f_{\sigma} \rangle m_{\rm *,disc}}{m_{\rm gas}}\right) \right]^{0.8}\,,
\end{equation}
we can reduce Equation (\ref{eq:model3}) to
\begin{equation}
\bar{\Sigma}_{\rm H\,{\LARGE{\textsc i}}}(\bar{r}) = \frac{\bar{\Sigma}_{\rm 0,H}\, \exp\!\big[\!-\!\bar{r} / \bar{r}_d \big]}{1 + \kappa\,\bar{\Sigma}_{\rm 0,H}^{1.6} \exp\!\big[\!-\!1.6\,\bar{r} / \bar{r}_d \big]}\,.
\label{eq:model3_red}
\end{equation}
By definition, it must hold true that $\bar{\Sigma}_{\rm H\,{\LARGE{\textsc i}}}\!=\!1$ when $\bar{r}\!=\!1$.  Therefore, it must also hold that
\begin{equation}
\kappa = \bar{\Sigma}_{\rm 0,H}^{-0.6}\, {\rm e}^{0.6 / \bar{r}_d} - \bar{\Sigma}_{\rm 0,H}^{-1.6}\, {\rm e}^{1.6 / \bar{r}_d}\,.
\end{equation}
We hence have a model with only two \emph{independent} parameters, as per our previous two cases.  The derived \HI~size--mass relation for this model is then
\begin{subequations}
\label{eq:m3sm}
\begin{multline}
\frac{m_{\rm H\,{\LARGE{\textsc i}}}}{r_{\rm H\,{\LARGE{\textsc i}}}^2} = 2\pi \int^{\infty}_0 \frac{\bar{r}\, \bar{\Sigma}_{\rm 0,H}\, {\rm e}^{-\bar{r}/\bar{r}_d}\, {\rm d}\bar{r}}{1 + \left( \bar{\Sigma}_{\rm 0,H}\,{\rm e}^{0.6/\bar{r}_d} - {\rm e}^{1.6/\bar{r}_d} \right) {\rm e}^{-1.6\,\bar{r}/\bar{r}_d}}\\
= 1.60769\, \pi\, \bar{\Sigma}_{\rm 0,H}\, \bar{r}_d^2~ _3\widetilde{F}_2 (a_1,a_2,a_3;~ b_1,b_2;~ c)\,,
\end{multline}
\begin{equation}
a_1 = a_3 = 0.625\,,~~a_2 = 1\,,
\end{equation}
\begin{equation}
b_1 = b_2 = 1.625\,,
\end{equation}
\begin{equation}
c = {\rm e}^{1.6/\bar{r}_d} - \bar{\Sigma}_{\rm 0,H}\,{\rm e}^{0.6/\bar{r}_d}\,,
\end{equation}
\end{subequations}
where $_3\widetilde{F}_2 (a_1,a_2,a_3;~ b_1,b_2;~ c)$ is the regularized hypergeometric function.
We note that \citet{wang14} previously identified that the Bluedisk galaxies' \SHIr~profiles are well fitted by an expression similar to Equation (\ref{eq:model3_red}): cf.~their equation 1.  The main differences here are that Equation (\ref{eq:model3_red}) (i) is derived from theory, rather than being empirically motivated, and (ii) has fewer free parameters.

Now we need to consider restrictions on the $(\bar{\Sigma}_{\rm 0,H}, \bar{r}_d)$ parameter space for model 3.  Firstly, the solution from Equation (\ref{eq:m3sm}) is only real when $c\!<\!1$.  This means we should uphold
\begin{equation}
\label{eq:ll}
\bar{r}_d > \left[ \ln\left(\bar{\Sigma}_{\rm 0,H}\right) \right]^{-1}\,.
\end{equation}
While we have already ensured that $\bar{\Sigma}_{\rm H\,{\LARGE{\textsc i}}}(\bar{r}\!=\!1)\!=\!1$, we should also ensure that ${\rm d}{\Sigma}_{\rm H\,{\LARGE{\textsc i}}} / {\rm d}\bar{r} |_{\bar{r}=1} < 0$ -- i.e.~the profile is declining at \rHI, not rising.  Enforcing this restricts the allowed sets of parameters further:
\begin{equation}
\label{eq:ul}
\bar{r}_d < - \left[ \ln\!\left( \frac{2.6 + \bar{\Sigma}_{\rm 0,H}^{0.6}}{1.6\,\bar{\Sigma}_{\rm 0,H} - \bar{\Sigma}_{\rm 0,H}^{0.4}} \right) \right]^{-1}\,.
\end{equation}
This right-hand side is only positive and finite for $\bar{\Sigma}_{\rm 0,H} \! \gtrsim \! 4.22$.  This provides a perfectly reasonable lower limit for the central surface density of neutral hydrogen in galaxy discs.  As we will show in the next subsection, in practice, observed galaxies only fill a very small area of this allowable parameter space, typically hugging the lower limit of Equation (\ref{eq:ll}).


\subsection{Comparison with observations}
\label{ssec:profiles}

\begin{figure}
\centering
\includegraphics[width=\textwidth]{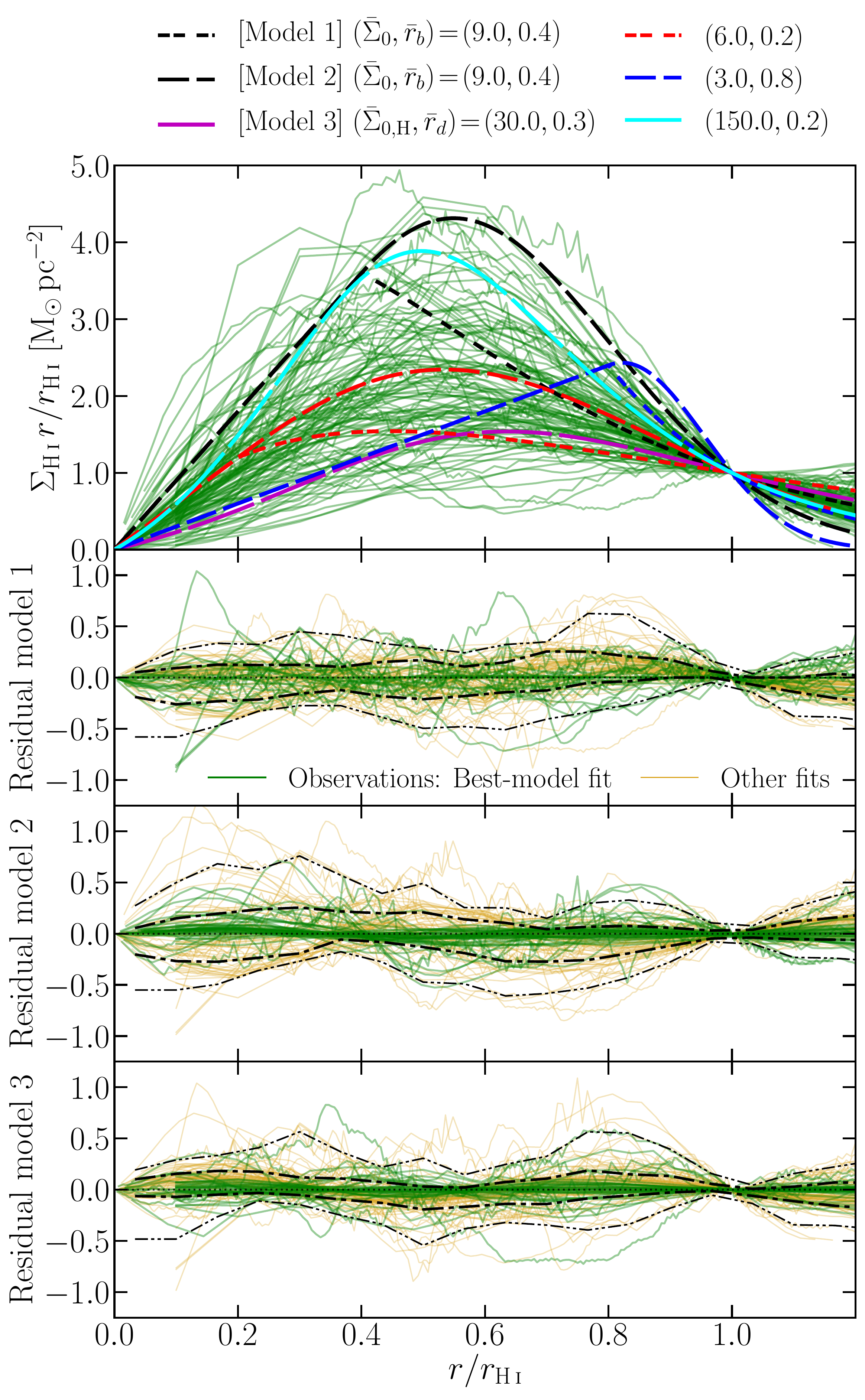}
\caption{Top panel: Normalized \HI~surface density integrand profiles for our full sample of observations (thin, solid curves), described in Section \ref{ssec:obs}; the area under each profile gives $m_{\rm H\,{\LARGE{\textsc{i}}}} / \left(2 \pi\, r_{\rm H\,{\LARGE{\textsc{i}}}}^2\right)$ for that galaxy (cf.~Equation \ref{eq:integral}).  For reference, overlaid are examples from our three analytic disc models, highlighting that the shape and variation of the model profiles (thicker, dashed curves) are qualitatively similar to observations.
Short dashes correspond to model 1 (Section \ref{ssec:model1}), medium-length dashes correspond to model 2 (Section \ref{ssec:model2}), and the longest dashes are for model 3 (Section \ref{ssec:model3}).  The colour of these thick curves represents a parameter set assumed for the model.  Three shorter panels: residuals for the best-fitting $\bar{\Sigma}_{\rm H\,{\LARGE{\textsc{i}}}}(\bar{r})\, \bar{r}$ profiles for each model to each observed galaxy.  Where residual profiles are green and more opaque, that model fit has the lowest $\chi^2$ of the three.  Thick, dot-dashed curves encompass 68 per cent of residuals, based on their interpolation onto a fixed $\bar{r}$ grid.  Thin, double-dot-dashed curves cover 95 per cent of residuals.}
\label{fig:profiles}
\end{figure}

In the top panel of Fig.~\ref{fig:profiles}, we show the \HI~surface density profiles for the sample of observed galaxies described in Section \ref{ssec:obs}.  Rather than showing \SHIr~by itself, we have multiplied the profiles by $r/r_{\rm H\,{\LARGE{\textsc{i}}}}$, as the area under these curves gives \mHI, and hence is what matters for the size--mass relation.  In other words, these are normalized integrand profiles.  A select few examples of analytic profiles from Equations (\ref{eq:SHIr}, \ref{eq:model2}, \& \ref{eq:model3}) are compared to help guide the eye, showing roughly that the available parameter space in each model covers the same area in the plot as the observed profiles, without going beyond.  To be more quantitative in this comparison, we have fitted  each individual profile with each analytic model.  In the lower panels of Fig.~\ref{fig:profiles}, we have subtracted the respective model best fits from each observed profile.  We overlay percentile ranges of the residuals from each model, and highlight the individual residuals in each panel where its corresponding model gives a better fit than the other two.  

In general, all three models capture the shape of the integrand profiles, with the area under the model curves closely shadowing those of the real profiles.  That the full two-sigma-equivalent residual range is at times nearly $1\,{\rm M}_{\odot}\,{\rm pc}^{-2}$ in height is not a cause for concern; the scatter here is driven by the fact that individual residuals oscillate about the zero line, meaning `bonus' area in parts of the profiles fits is typically cancelled by `missing' area in other parts of the same fit.  Indeed, some of the residuals show a significant amount of noise, owing to the simplicity of the fits and the lack of consideration of disc asymmetries (in principle, the observed profiles should have projection effects accounted for).  While we do not suggest that there is a clear `best' model, we note that the scatter in the residuals is marginally smaller for model 3 than the others ($\sim\!0.11$ versus $\sim\!0.14\,{\rm M}_{\odot}\,{\rm pc}^{-2}$), but model 2 provides the best fit the most often (46 times versus 28 and 36 for models 1 and 3, respectively).

\begin{figure*}
\centering
\includegraphics[width=\textwidth]{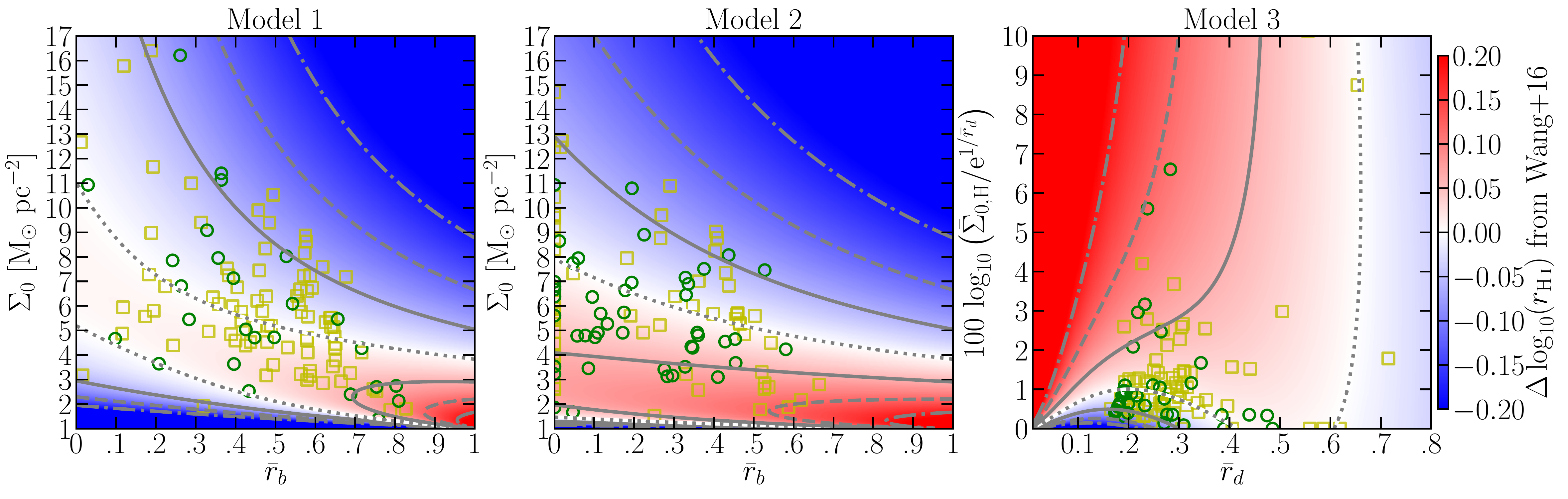}
\caption{Maps of how far scattered from the best-fitting, observed \HI~size--mass relation galaxies would be, based on their location in parameter space for each of our three analytic gas disc models.  Solid, dashed, and dot-dashed contours represent where the galaxies lie $1\sigma$, $2\sigma$, and $3\sigma$ from the \citetalias{wang16} relation, respectively (where $\sigma\!=\!0.06\,{\rm dex})$.  The dotted contour represents a displacement of zero.  Square and circles represent the best-fitting parameters to observed \HI~profiles; circles indicate that that model gives a better fit than the other two for that particular galaxy.}
\label{fig:scattermap}
\end{figure*}

Subjectively more interesting than the scatter in the residuals of \HI~profiles is the scatter in the \HI~size--mass relation if one were to take each or any of these models as representative of galaxies in the real Universe.  The size--mass relations derived from all three models predict a slope of 0.5.  In order for there to be a direct mapping for a parameter pair to a displacement from the real \HI~size--mass relation, the real relation would also need to have the same slope (otherwise we would need to introduce a tertiary mass dependence).  Given how close the slope measured by \citetalias{wang16} is to 0.5, we assume for the purposes of calculations throughout this paper (with the exception of Section \ref{ssec:infer}) that 0.5 is indeed the true slope (rather than 0.506).%
\footnote{Other works with different galaxy samples have found slopes slightly more deviant from 0.5 (or 2.0, dependent on axis orientation) than this (e.g.~\citealt{lelli16}; \citealt*{ponomareva16}).}  
The intercept in the relation also requires minor modification to reflect this.  We choose to preserve the \HI~size of galaxies exactly at an \HI~mass of $10^9\,{\rm M}_{\odot}$ (the typical mass for the observations and the simulations we use later).  We therefore treat the observed intercept as $-3.239$ (rather than $-3.293$).  

With these assumptions in place, we show maps of how far scattered galaxies would be from the observed \HI~size--mass relation based on their parameters for each of our three models in Fig.~\ref{fig:scattermap}.  To help navigate these maps, contours highlight where the scatter values correspond to integer numbers of standard deviations.  Overlaid on these plots, we show the best-fitting parameter values from our observational sample.  For models 1 and 2, we show the full range of allowable values of $\bar{r}_b$ and extend the range in $\Sigma_0$ out to $17\,{\rm M}_{\odot}\,{\rm pc}^{-2}$; while we do not expect an abundance of galaxies to have such a high value of $\Sigma_0$, some of the fits to observations almost reach this.  For model 3, the observations guide the area of parameter space that we plot.  This necessitated reframing the way the parameter space is visualized -- i.e.~not just $\bar{r}_d$ versus $\bar{\Sigma}_{\rm 0,H}$, as these properties are highly (anti-)correlated.  Per Equation (\ref{eq:ll}), we know $\bar{\Sigma}_{\rm 0,H}\,\exp(-\bar{r}_d^{-1}) \! > \! 1$ always.  As it happens, the fits to all the observations find values no higher than 1.3 for this quantity.

The main message of Fig.~\ref{fig:scattermap} is that effectively any galaxy that follows any of our three models -- with parameters in a physically plausible and meaningful range -- will be consistent with the observed \HI~size--mass relation.  As one would expect for a sample size of $\sim\!100$, most of the observational points fall between the $\pm1\sigma$ contours, with a small number approaching $\pm2\sigma$, and only a hint that the odd galaxy would lie further away.  Given that these models generally fit the observations well, and that the distributions of parameters associated with those fits are consistent with nominal expectation (see Table \ref{tab:params}), a tight relation between \HI~size and \HI~mass is arguably a simple inevitability.

\subsubsection{Inferred model size--mass relations}
\label{ssec:infer}

In order to get the actual normalization, scatter, and slope of the predicted \HI~size--mass relation for each model, one needs to know how the parameter space of each model is occupied (and whether there are any implicit mass biases for parts of the parameter space).  It is unclear \emph{a priori} what the distribution functions of these parameter spaces should be.  What we can do, though, is use the parameter fits to the observed profiles in our sample, and assume that this sample is representative of the underlying parameter space distributions.  While this assumption is not robust (see Section \ref{ssec:obs}), it should be sufficiently accurate for us to make a relative comparison of the derived size--mass relations from each model.
In practice, this means fitting the relations to the observed sample in several ways, where \rHI~remains the same for a given galaxy in all cases, and all that changes for the different models is that the empirical \mHI~is replaced by the analytic value derived from the parameter fits.

\begin{figure}
\centering
\includegraphics[width=\textwidth]{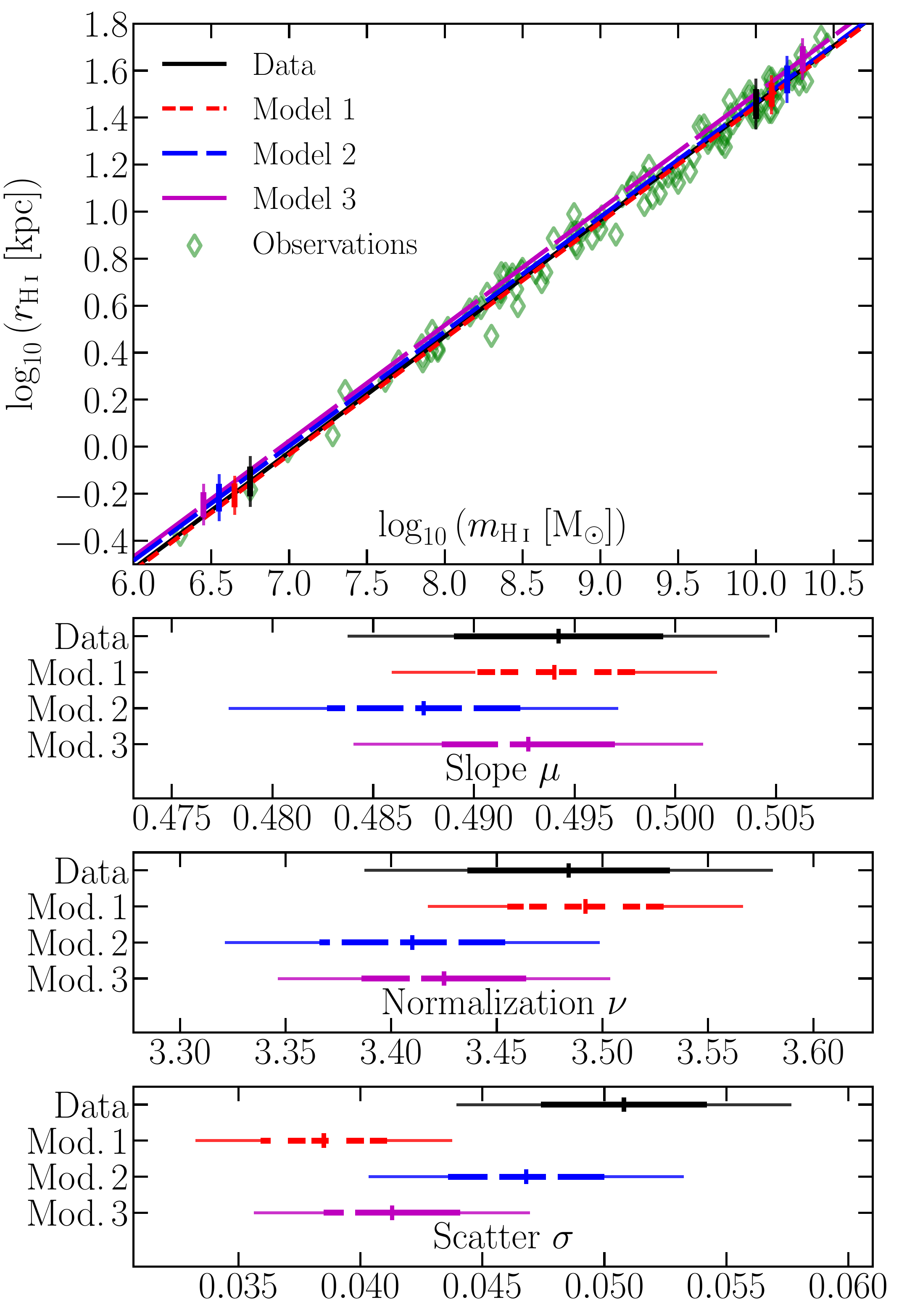}
\caption{Top panel: \HI~size--mass data and fitted relation for our sample of observations (Section \ref{ssec:obs}); diamonds are individual galaxies, and the solid line labelled `data' gives the best fit to these data.  The other lines are the predicted \HI~size--mass relations for each of our analytic models, assuming their parameter spaces to be occupied consistently with the \HI~profile fits to the observations.  Vertical bars show the 1$\sigma$ (thick) and 2$\sigma$ (thin) scatter in each relation. Bottom 3 panels: parameters for the \HI~size--mass relation fits.  Vertical ticks show the best-fitting values, assumed in the top panel.  Horizontal bars show the uncertainty ranges for each parameter (thick for one standard deviation, thin for two).  These are listed in Table \ref{tab:modrat}.}
\label{fig:HISM}
\end{figure}

In Table \ref{tab:modrat}, we collate the \HI~size--mass relations for each model, derived with the above method.  To fit the size--mass relations, we use the {\sc hyper-fit} \citep{robo15} web interface\footnote{\url{http://hyperfit.icrar.org/}} with default settings.  {\sc hyper-fit} uses a Bayesian approach to find the maximum likelihood of a linear model that describes multidimensional data.  We ignored any uncertainties on the individual data when making the fits.  We fit and include in Table \ref{tab:modrat} the size--mass relation using the `true' \HI~masses of the galaxies too (from numerically integrating their observed surface density profiles).  As one would expect, this fit differs from \citetalias{wang16} because (i) our sample is only a subset of theirs, (ii) the code to make the fit is not the same, and (iii) our \mHI~measurements for the Bluedisk galaxies differs.  All these \HI~size--mass relations and the parameter ranges are plotted in Fig.~\ref{fig:HISM}.  The slope, scatter, and normalization of all the relations each overlap within $\lesssim\!2$ standard deviations of their {\sc hyper-fit} Gaussian uncertainties.  

\begin{table*}
\centering
\begin{tabular}{l c c c c} \hline\hline
& Data & Model 1 & Model 2 & Model 3 \\ \hline\hline
Slope $\mu$ & $0.4942\pm0.0052$ & $0.4940\pm0.0040$ & $0.4875\pm0.0048$ & $0.4927\pm0.0043$ \\ 
Normalization $\nu$ & $3.484\pm0.048$ & $3.492\pm0.037$ & $3.410\pm0.044$ & $3.425\pm0.039$ \\
Scatter $\sigma$ & $0.0508\pm0.0034$ & $0.0385\pm0.0026$ & $0.0468\pm0.0032$ & $0.0413\pm0.0028$ \\ \hline\hline
\end{tabular}
\caption{\HI~size--mass relation fits to our sample of observational data (Section \ref{ssec:obs}), where $\log_{10}\!\left(r_{\rm H\,{\LARGE{\textsc i}}} / {\rm kpc} \right) \!=\! \mu\log_{10}\!\left(m_{\rm H\,{\LARGE{\textsc i}}} / {\rm M}_{\odot}\right) - \nu \pm \sigma$.  All fits have been made with {\sc hyper-fit} \citep{robo15}.  The `data' column is a direct fit to the observed \rHI~and \mHI~values.  The `model' columns use the \mHI~given by the best-fitting model \HI~profile for each galaxy.}
\label{tab:modrat}
\end{table*}

The nominal conclusion we draw from this exercise is that all our analytic models predict \HI~size--mass relations that are not just qualitatively, but also quantitatively consistent with observations.  We should stress that this conclusion has been reached imperfectly though; ideally the distributions of the model parameter spaces should be derived or explored independently from the data we compare to.  This is left as a task for future work.  In the meantime, more information on the model parameter distributions is given in Appendix \ref{app:dists}.

\subsection{Comparison with simulations}
\label{ssec:sims}

For context, before addressing how well our analytic \HI~surface density profiles are reflected in cosmological simulations, we should first address how well those simulations reproduce the observed \HI~size--mass relation.  
Recently, \citet[][see their fig.~5]{diemer19} showed that the \HI~size--mass relation of TNG100 (and TNG300) galaxies at \zo~follows that of \citetalias{wang16} but for a small systematic offset and a slightly larger scatter.
Similarly, \citet[][see their fig.~3]{lutz18} previously showed that the original version of \ds~\citep{stevens16} reproduced the observed \HI~size--mass relation, almost precisely matching \citetalias{wang16} but for a smaller scatter.
Because the \HI~structure of galaxies is grown numerically in both TNG and \ds, and this structure is subject to a large number of astrophysical processes relevant for galaxy evolution, these simulations provide a far more comprehensive tool for predicting and analysing the \HI~size--mass relation than simple analytic models.  
To summarize their relations (and update in the case of \ds), we provide their normalizations and scatters in Table \ref{tab:relations}.  We obtained the normalizations with a least-squares linear fit in log-log space, assuming a slope of 0.5 (in accordance with the analytic predictions).  The scatter values are then standard deviations of the residuals between the fitted and actual \HI~sizes of the galaxies.  As per Sections \ref{ssec:tng} and \ref{ssec:ds}, for both TNG100 and \ds, we only consider resolved galaxies with $m_* \! \geq \! 10^9\,{\rm M}_{\odot}$ and $m_{\rm H\,{\LARGE{\textsc i}}} \! \geq \! 10^8\,{\rm M}_{\odot}$.

\begin{table}
\centering
\begin{tabular}{l c c} \hline
Data source & $\nu$ & $\sigma$ \\ \hline
Observations \citepalias{wang16} & 3.540 & 0.060\\
TNG100 (Section \ref{ssec:tng}) & 3.516 & 0.095\\
\ds~(Section \ref{ssec:ds}) & 3.603 & 0.051 \\ \hline

\end{tabular}
\caption{The normalization and scatter (standard deviation) of the best-fitting \HI~size--mass relations from observations and our simulations.  All assume a fixed-slope relation of $\log_{10}\!\left(r_{\rm H\,{\LARGE{\textsc i}}} / {\rm kpc} \right) \!=\! 0.5\log_{10}\!\left(m_{\rm H\,{\LARGE{\textsc i}}} / {\rm M}_{\odot}\right) - \nu \pm \sigma$.  The values for observations are taken initially from \citet{wang16}, but re Section \ref{ssec:profiles}, the normalization has been modified to match the assumption that the slope is 0.5.  The standard deviation quoted for simulations is cleaned for outliers; an initial standard deviation, $\sigma_{\rm all},$ is first calculated for all galaxies, then $\sigma$ is recalculated after removing galaxies lying at $>\!3\sigma_{\rm all}$.  Both \ds~and TNG100 had $\sigma_{\rm all} \! > \! 0.11$.}
\label{tab:relations}
\end{table}

\begin{figure*}
\centering
\includegraphics[width=0.85\textwidth]{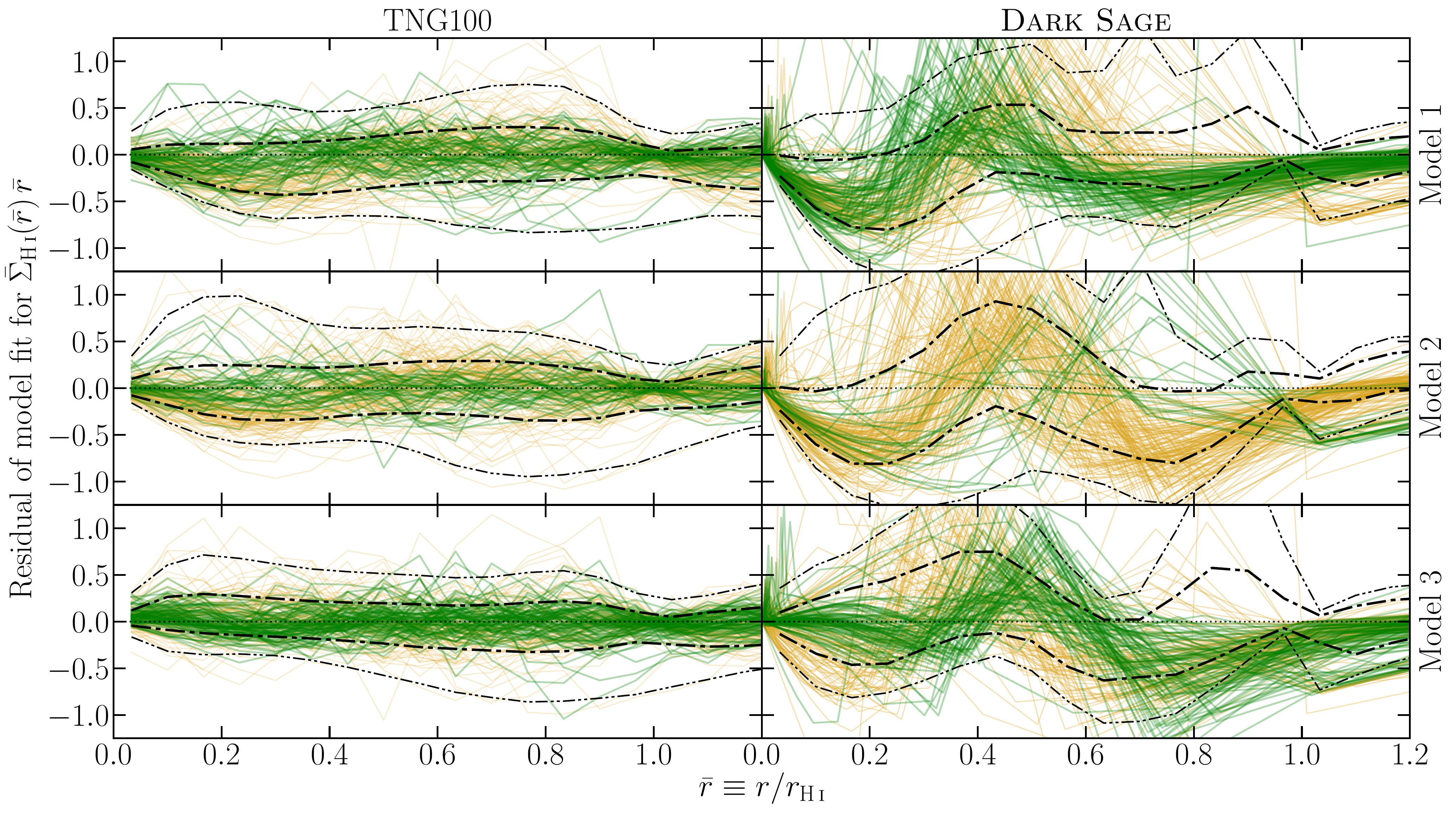}
\caption{Residuals to analytic fits for our model profiles to $\sim$200 example galaxies with $m_{\rm H\,{\LARGE{\textsc i}}} \! \geq \! 10^9\,{\rm M}_{\odot}$ each from TNG100 and {\sc Dark Sage}.  Running percentiles use the full samples (with $m_{\rm H\,{\LARGE{\textsc i}}} \! \geq \! 10^8\,{\rm M}_{\odot}$) and are built on a common grid.  Plotting convention matches that of the bottom three panels of Fig.~\ref{fig:profiles}.  The individual \ds~residuals follow their proper annular profiles, where the spacing of the annuli increases exponentially with radius; the combination of this with \rHI~being fixed in the fits leads to excessive noisiness in the residuals around \rHI. Further details are described in Section \ref{ssec:sims}.}
\label{fig:sim_residuals}
\end{figure*}

As with the observations, to see how well our analytic \HI~profiles reflect those predicted by the simulations, we fit each simulated galaxy with each model.  For TNG100 galaxies, we build one-dimensional \HI~surface density profiles on a radial fixed grid of bin width $\frac{2}{30}$\,\rHI~out to 1.6\,\rHI.  Each analytic model is fitted to the integrand $\bar{\Sigma}(\bar{r})\,\bar{r}$ profiles using a $\chi^2$ minimization.  In the left panels of Fig.~\ref{fig:sim_residuals}, we show residuals for these fits for $\sim$200 randomly selected TNG100 galaxies with $m_{\rm H\,{\LARGE{\textsc i}}} \! \geq \! 10^9\,{\rm M}_{\odot}$, along with running percentiles for the full sample.  \ds~already has defined bins within which \SHI~is produced for each galaxy.  However, because they increase in width exponentially with radius, fitting to these bins as is would be ineffective (i.e.~often non-convergent), as this would weight the entire fit to the galaxy centre, where the contribution to overall \HI~mass is minimal.  We therefore instead interpolate the inherent \SHIr~profiles onto the same radial grid used for TNG100, then fit each model to $\bar{\Sigma}(\bar{r})\,\bar{r}$ on that grid.  The right panels of Fig.~\ref{fig:sim_residuals} give examples and running percentiles of the residuals for the full \ds~sample.  

The \HI~profile fits to neither TNG100 nor \ds~are as close as they were for the observed sample; the typical scatter in the residuals is a factor of $\sim$2 and $\sim$3 larger, respectively.  Nevertheless, all three analytic profiles generally reflect the shape of TNG100 profiles, with model 3 edging model 2 for the lowest scatter in the residuals, and models 3 and 1 each giving twice the number of lowest-$\chi^2$ fits than model 2.  For \ds, model 3 most often gives the best fit, but the overall scatter in the model-1 fits is lower.  At some level, the noisiness of the \ds~residuals cannot be helped by the way the discs are pre-constructed with discrete annuli.  What these plots hide is that the \HI~mass returned by passing the fitted parameter values back through the model equations (using the true \rHI) are more faithful to the true values for \ds~than they are for TNG100.  And for both simulations, the returned \mHI~values for the model-3 fits are the least faithful, while those from model 1 are the most accurate.
See Appendix \ref{app:dists} for an overview of the fitted profile parameters to both simulations.


\subsubsection{Variation with galaxy type}
\label{ssec:type}

\begin{figure}
\centering
\includegraphics[width=\textwidth]{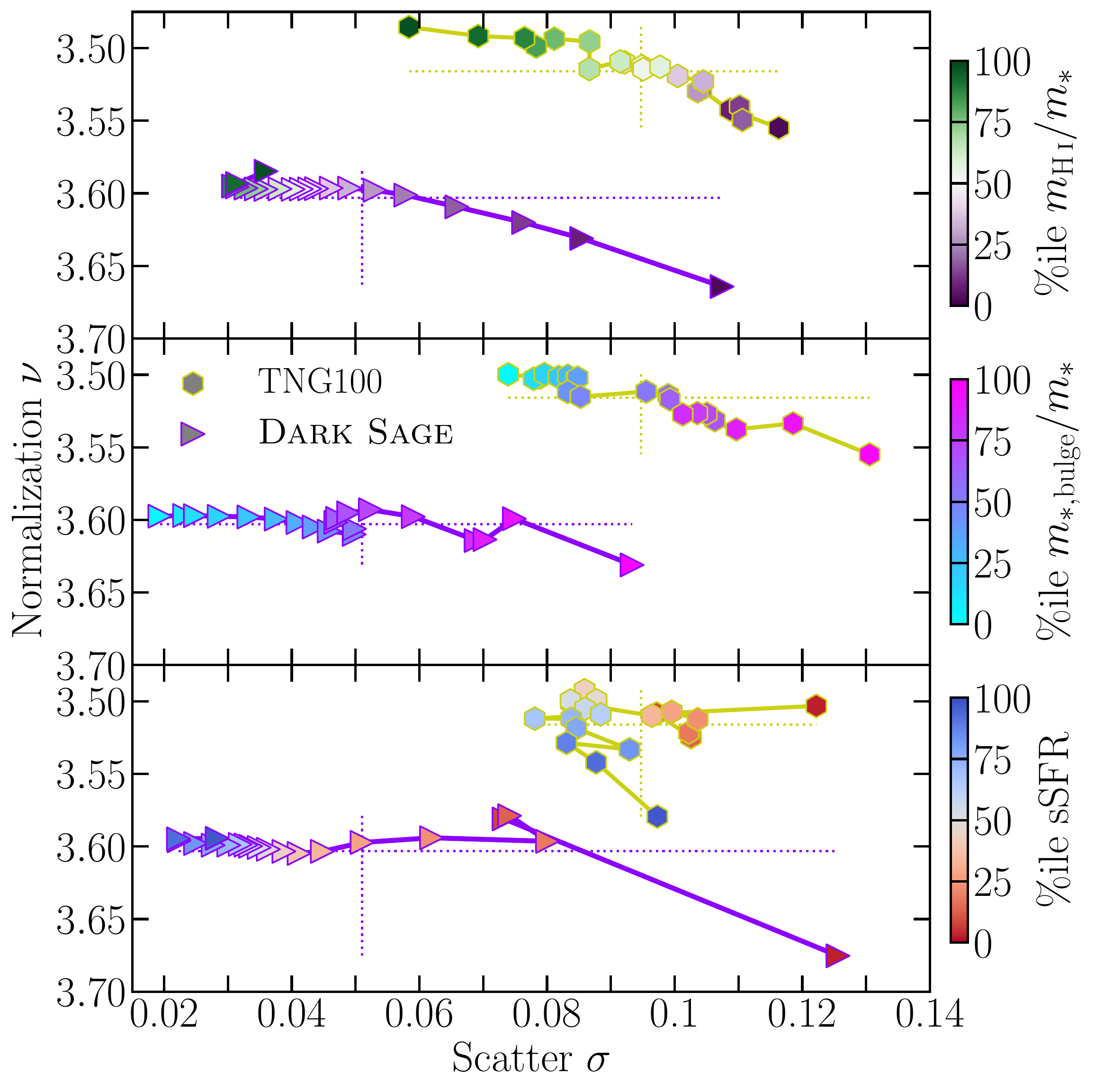}
\caption{Variation in the best-fitting scatter and normalization of the \HI~size--mass relation (of fixed slope $\mu\!=\!0.5$) for TNG100 and \ds~galaxies when selecting on \HI~fraction (top panel), bulge-to-total ratio (middle panel), and specific star formation rate (bottom panel) for fixed percentile ranges.  In general, the more quenched, bulge-dominated, and/or \HI-poor a population of galaxies is, the lower the average \HI~size and wider the distribution of \HI~sizes of that population at fixed \HI~mass.  Horizontal and vertical dashed lines intersect at the values for the full simulation samples (given in Table \ref{tab:relations}).}
\label{fig:type}
\end{figure}

An outstanding question surrounding the \HI~size--mass relation is whether it is equally applicable to galaxies of all types.  That is, do quenched/bulge-dominated/dispersion-supported/gas-poor galaxies have a common \HI~size--mass relation with star-forming/disc-dominated/rotation-supported/gas-rich galaxies?  Observational studies have typically lacked a sufficiently large and simultaneously diverse enough sample of galaxies to address this directly. Where we can more readily find insight is from our sample of simulated galaxies.  
To achieve this, we rank order our TNG100 and \ds~galaxies in three ways: (i) by their \HI-to-stellar mass ratio, (ii) by their stellar bulge-to-total mass ratio, and (iii) by specific star formation rate (${\rm sSFR} \! = \! {\rm SFR}/m_*$).  Then we refit the \HI~size--mass relation for bins in each property of fixed galaxy number, maintaining an assumed slope of 0.5.  In Fig.~\ref{fig:type}, we show how the normalization and scatter of these fits vary.  By binning galaxies this way, rather than on absolute values of the same properties, we avoid caveats surrounding systematic differences in galaxy properties between the simulations and how properties like bulge mass are defined.%
\footnote{Nevertheless, for completion, we note that SFRs for TNG100 galaxies are calculated from the instantaneous rates of the gas cells, while \ds~uses time-averaged quantities across the previous snapshot interval in the Millennium merger trees.  \ds~bulges include contributions from mergers and instabilities but \emph{not} the pseudobulge (see \citealt{stevens16,stevens18} for clarification).  TNG100 stellar particles are classed as being in a rotationally supported disc if they fulfil the criteria
$\left\lvert \log_{10}\left(\frac{2\, \mathcal{K}_{\rm tangential}}{\mathcal{U}_{\rm gravitational}}\right) \right\rvert < 0.2$
and
$\mathcal{K}_{\rm tangential} > 2\, \mathcal{K}_{\rm radial}$
\citep[similar to][where $\mathcal{K}$ and $\mathcal{U}$ are kinetic and potential energy per unit mass, respectively]{mitchell18}.  The remaining stellar particles make up the bulge.  This approach for TNG100 is sufficient for defining \emph{relative} morphologies, but not absolute.}

For all 3 galaxy property variations, \ds~finds a robust normalization to the \HI~size--relation that only decreases when one selects galaxies with the least gas/star formation activity and/or the biggest bulge fraction \citep[cf.][]{lutz18}.  But the scatter steadily increases as one moves towards that end of the spectrum, with a difference of a factor of $\sim$6 between the two extremes.  Although, even for $\sigma\!=\!0.125\,{\rm dex}$ (a scatter of 33\%), the relation is still objectively tight by astrophysical standards.  TNG100 exhibits similar behaviour when selecting on \HI~fraction or bulge fraction, but also shows a steady decline in normalization.  When selecting on sSFR, the situation is less ordered for TNG100.  While this result highlights that the preciseness of a derived \HI~size--mass relation is dependent on the underlying galaxy sample (i.e. whether it is representative or biased), variations in the normalization are generally smaller than the relation's scatter.  No galaxy selected on the properties in Fig.~\ref{fig:type} would therefore look like an outlier from the representative \HI~size--mass relation.

In summary, while the \SHIr~profiles of galaxies produced by cosmological simulations are not all \emph{precisely} characterized by a common analytic form, their behaviour is similar enough to the three models presented in this Section, such that all methods are ultimately consistent in predicting a tight \HI~size--mass relation with minimal wiggle room in its slope, normalization, and scatter, in the absence of heavy biases.


\section{Environmental stripping of gas}
\label{sec:env}

In this section, we assess one potential method for disrupting the \HI~profiles of galaxies, that being the environmental stripping of gas.  Taking the analytic profiles proposed in Section \ref{sec:models} as a starting point, we make analytic predictions for how disc truncation might impact the \HI~size--mass relation, if at all.  While we motivate ram-pressure stripping as a mechanism for disc truncation, the following is agnostic to the motivation.  Tidal stripping, for example, can also contribute to the truncation of a disc.  We do not assess how the induced asymmetries from tides or ram pressure (the leading side of the galaxy should experience greater pressure, e.g. \citealt{chung09}) might fold into the \HI~size--mass relation.  We use results from TNG100 and \ds~as a means of testing and expanding on our analytic work; both simulations have far more complete considerations of galaxy environment (implicitly and explicitly, respectively).  Unfortunately, we have too few and insufficiently diverse observational data to check this against the real Universe.

\subsection{Disc truncation}
\label{ssec:truncation}

When accounting for ram pressure on a cold-gas disc, gas is typically regarded as being stripped below the threshold surface density where the gravitational restoring force per unit area is insufficient to counterbalance the ram pressure \citep{gunn72}.  Assuming that the strength of gravitational restoration falls off with disc radius (which is a given for gas disc profiles whose gradients are negative or nil everywhere, true for all models considered in Section \ref{sec:models}), the \HI~profiles of satellites experiencing ram pressure should become progressively \emph{truncated} with time.  Indeed, ram pressure has been implemented in several semi-analytic models of galaxy formation this way \citep[][]{lanzoni05,tecce10,luo16,stevens16}.  

Assuming any of Equations (\ref{eq:SHIr}, \ref{eq:model2}, or \ref{eq:model3_red}), we can analytically show how disc truncation would affect the \HI~size--mass relation of galaxies.  
To find \mHI~for a galaxy with a truncated disc, we simply need to integrate \SHIr\,$r$ out to the truncation radius, $r_t$.  \rHI~will only change from its initial value (hereafter denote as $r_{\rm H\,{\LARGE{\textsc i}}, init}$) if $r_t$ is smaller than it.
That is, $r_{\rm H\,{\LARGE{\textsc i}}} \rightarrow {\rm min}\left(r_t, r_{\rm H\,{\LARGE{\textsc i}}, init}\right)$.
The explicit equations for all three model profiles undergoing truncation are provided in Appendix \ref{app:trunceq}.
Using these, in Fig.~\ref{fig:trunc}, we show tracks for how galaxies would move in the \HI~size--mass plane as they are truncated to continually smaller radii.  

\begin{figure}
\centering
\includegraphics[width=\textwidth]{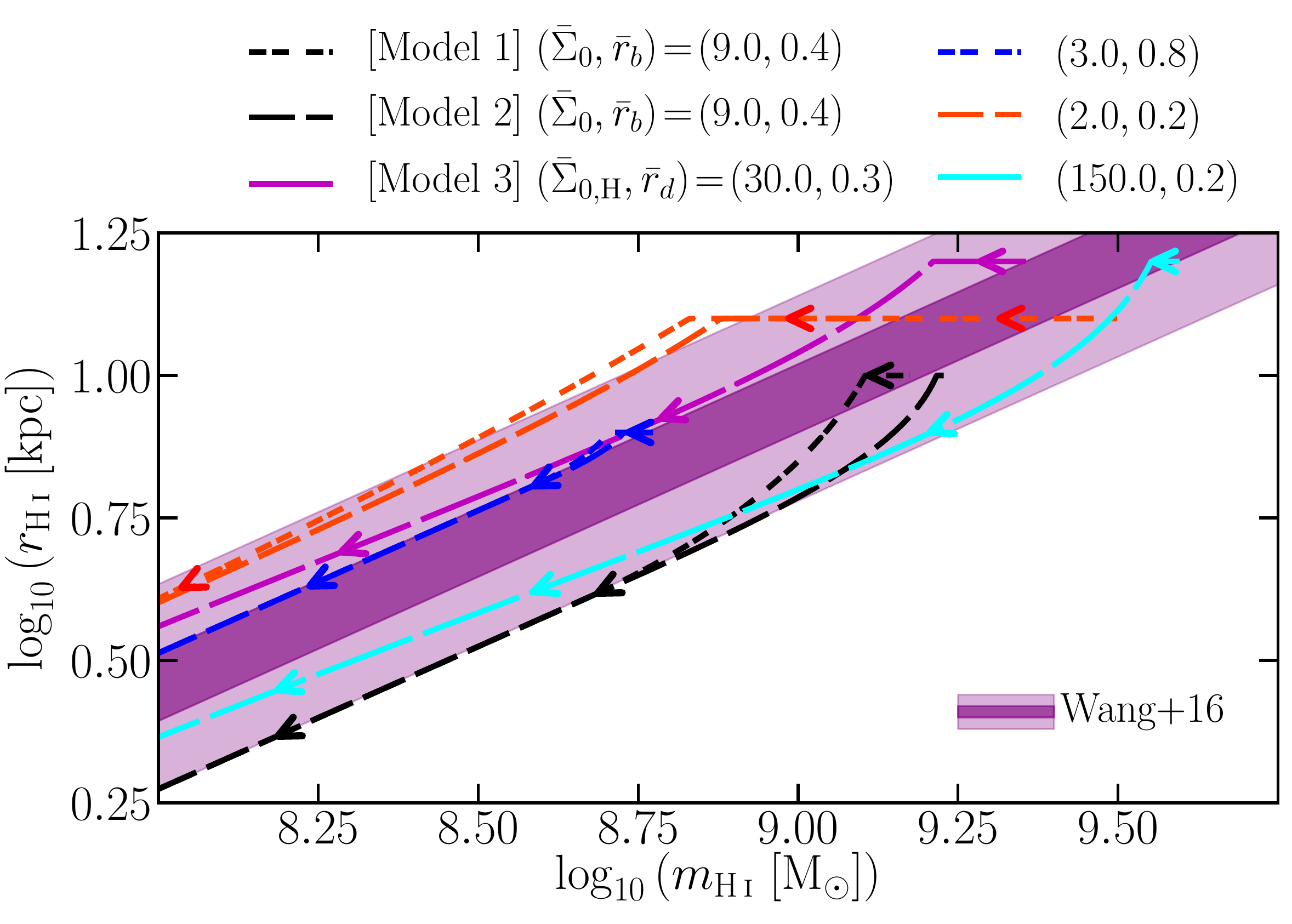}
\caption{Tracks for example galaxies in the size--mass plane when their \HI~discs become progressively truncated.  Galaxies start in the top right of each track, following the direction of the arrows, having been manually assigned an initial \rHI.  The precise path depends on whether the discs are initialized assuming model 1 (short dashes), model 2 (medium-length dashes), or model 3 (longest dashes).  Line colour differentiates parameter choices for the models.  Equations are provided in Appendix \ref{app:trunceq}.}
\label{fig:trunc}
\end{figure}

For models 1 and 2, there are three phases seen in each track in Fig.~\ref{fig:trunc}.  Starting from the top right, first is the horizontal part of the track, where $\bar{r}_t \! > \! 1$ and thus \mHI~reduces even though \rHI~remains the same.  The second part of the track is (the only part that is) curved and concave up, where $\bar{r}_b \leq \! \bar{r}_t \! \leq \! 1$.  The third, diagonally straight part of the track covers $\bar{r}_t \! < \! \bar{r}_b$. While the displacement of the galaxies in the size--mass plane from the best-fitting relation of \citetalias{wang16} changes during truncation -- with some being scattered up/left, some scattered down, and some returning to their original displacement -- the galaxies still remain generally within the observed scatter as a natural consequence of the equations governing the tracks.

The truncation tracks for model 3 are qualitatively similar to the other models but with some subtle differences.  Naturally, they all share the same initial horizontal path where $\bar{r}_t \! > \! 1$.  The tracks then have similar curvature for $\bar{r}_t \! \leq \! 1$.  But rather than reaching a point where the gradient becomes fixed, it instead continues to decrease (moving from right to left), going below 0.5, before becoming concave down and reapproaching 0.5 asymptotically.  As such, they also do not diverge from the observed size--mass relation.

The conclusion we draw is that galaxies undergoing environmental stripping are generally not outliers in the \HI~size--mass relation.  What is more, this is not necessarily restricted by our decision to model gas stripping as the progressive truncation of a satellite's disc.  To back that up, let us consider now that ram pressure (or any environmental process) not only leads to truncation, but also to an overall suppression of gas surface density \citep[see e.g.][]{cayatte94}.  For models 1 and 2, if \SHIr~drops by a uniform fraction across the disc, then $\Sigma_0$ drops and $\bar{r}_b$ increases (as \rHI~decreases but $r_b$ does not).  Similarly for model 3, $\bar{\Sigma}_{\rm 0,H}$ would drop and $\bar{r}_d$ would rise.  In all cases, the galaxy would still reside within the region of the respective model's parameter space assessed above, and therefore the galaxy would still conform to the observed size--mass scatter.  Furthermore, any change in $r_s$ could simply be captured as a change in $\bar{r}_b$ or $\bar{r}_d$. The only way a galaxy would become an outlier in the \HI~size--mass relation is for the functional form of its \SHIr~profile to undergo a drastic change such that it no longer resembles any of Equations (\ref{eq:SHIr}, \ref{eq:model2}, or \ref{eq:model3}).


\subsection{Results from D{\small ARK} S{\small AGE}}
\label{ssec:truncDS}

Let us now examine what effect galaxy environment has on the \HI~size--mass relation in the \ds~semi-analytic model.  \ds~provides a trustworthy and logical numerical experiment to test the picture described in Section \ref{ssec:truncation} for two main reasons.  First, \citet{sb17} have already shown how the model predicts that environment impacts galaxies' \HI~content similarly to what is observed at \zo~\citep[also see][]{stevens18}.  Secondly, cold-gas stripping is explicitly implemented in the model by finding the innermost annulus of a satellite galaxy's disc where there is insufficient restoration from gravity to balance the ram pressure it experiences as it travels through its parent halo's hot gas medium, and truncates the disc there.  Because \ds~is run on a $500\,h^{-1}\,{\rm Mpc}$ box, there is plenty of statistical power in galaxies across all environments.

\begin{figure}
\centering
\includegraphics[width=\textwidth]{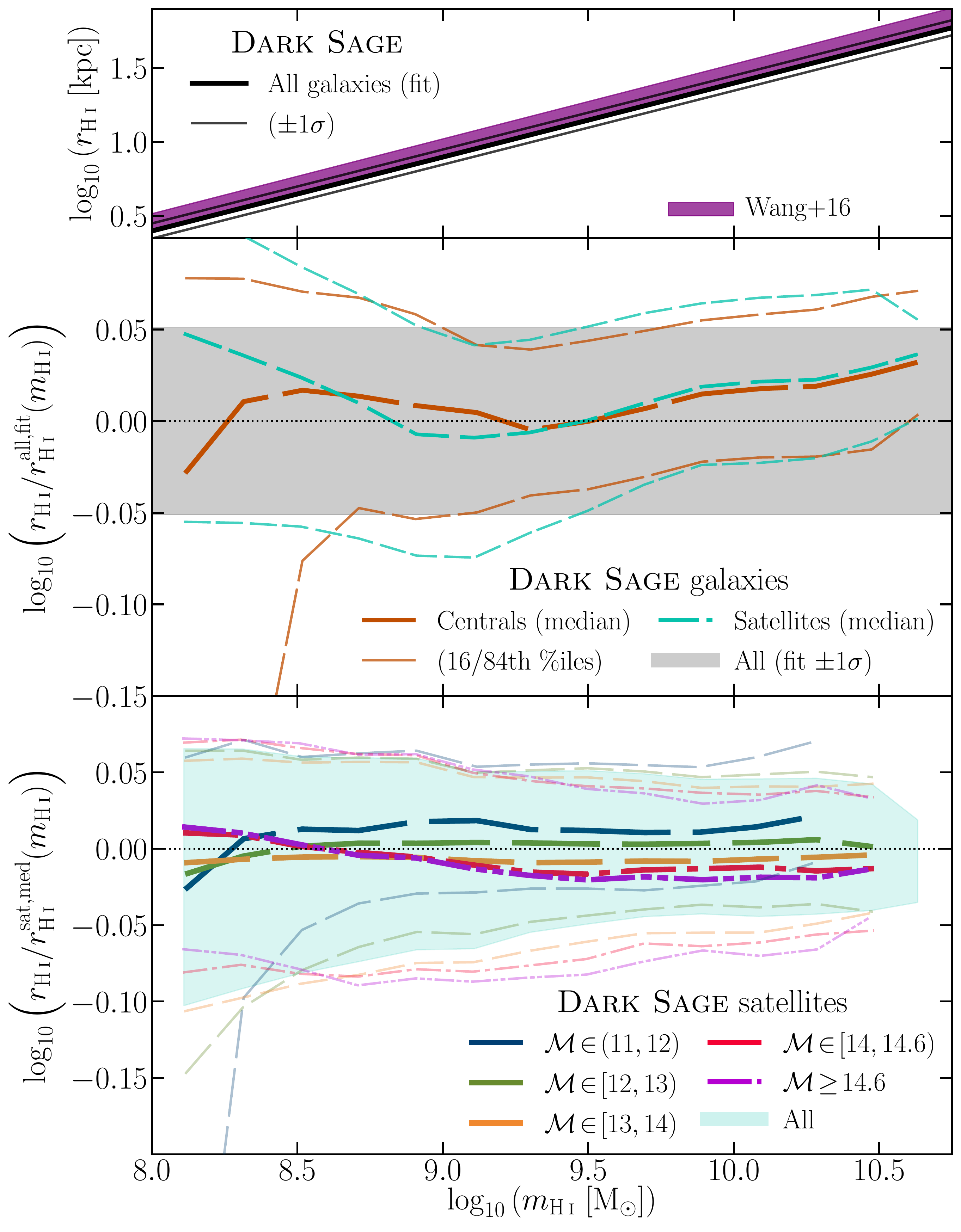}
\caption{Top panel: Best-fitting \HI~size--mass relation for all \ds~\citep{stevens18} galaxies at \zo~with $m_*\!\geq\!10^9\,{\rm M}_{\odot}$ and $m_{\rm H\,{\LARGE{\textsc i}}}\!\geq\!10^8\,{\rm M}_{\odot}$ (see Table \ref{tab:relations}).  This assumes a slope of 0.5, and is compared to the observed relation of \citetalias{wang16}.  The $1\sigma$ scatter in both relations is shown.
Second panel: \emph{Difference} in the \HI~size of \ds~central and satellite galaxies relative to the fitted relation in the top panel.  Running medians (thick curves) and percentiles (thin curves) are given for both galaxy types (differentiated by dash style and colour).  The grey shaded region covers $\pm$\,one standard deviation from the fitted relation.
The bottom panel compares the difference in \HI~size for satellites in denoted halo mass bins [$\mathcal{M} \!\equiv\! \log_{10}\left(M_{\rm 200c}/{\rm M}_{\odot}\right)$] \emph{to the median for all satellites} at the same \HI~mass.  Thick and thin lines still refer to the median and 16th/84th percentiles here, respectively.  Longer dashes in the lines correspond to lower halo masses.  The lightly shaded region in the bottom panel covers the 16th--84th percentile range for all satellites (the same as the sandwiched range for satellites in the second panel, provided for reference along with the horizontal dotted line at 0).  All percentiles for all panels are calculated in bins of minimum width 0.2\,dex in $\log_{10}(m_{\rm H\,{\LARGE{\textsc i}}})$, each with a minimum of 20 galaxies.}
\label{fig:SizeMassDS}
\end{figure}

In the top panel of Fig.~\ref{fig:SizeMassDS}, we show the best-fitting \HI~size--mass relation for the \citet{stevens18} version of \ds.  This assumed a fixed slope of 0.5 (the normalization and scatter are given in Table \ref{tab:relations}).  We then break galaxies into centrals and satellites in the middle panel, showing deviations (or lack thereof) from the fitted relation for all galaxies on the $y$-axis.  The distinction between satellite and central provides a zeroth-order consideration of environment, as only satellites are subject to stripping processes (by construction, as described in Section \ref{ssec:ds}).  Almost no difference is seen between centrals and satellites; only towards the resolution limit ($m_{\rm H\,{\LARGE{\textsc i}}}\!\lesssim\!10^{8.5}\,{\rm M}_{\odot}$) does anything become apparent, and that should not be overanalysed.  This is in contrast to their difference in \HI~mass at fixed stellar mass, for example \citep[see fig.~3 of][]{sb17}.  In fact, over most of the considered mass range, the median lines for centrals and satellites both run close to the zero line (i.e.~in line with the fitted relation for all galaxies), as do the 16th and 84th percentiles for both only deviate moderately from the edges of the $\pm1\sigma$ range of the fit.  Already this tells us that environment does not have more than a secondary effect on \HI~size--mass relation (if any), consistent with the derivations in Section \ref{ssec:truncation}.

To properly test this, we need to more quantitatively break galaxies into different environments.  This mandates that we define a metric for environment.  We choose to use the parent halo mass of a galaxy for this.  Observations suggest this is a more meaningful metric than, for example, galaxy number density based on the $N$th nearest neighbour \citep[e.g.][]{brown17}.  The greater the halo mass, the denser the typical intrahalo gas medium the satellites will move through, and the faster they will move through it. Therefore, the effects of stripping will be stronger on the satellites.  This is demonstrably true for \ds~\citep[surmisable from][]{stevens16,stevens18,sb17}.

The bottom panel of Fig.~\ref{fig:SizeMassDS} dissects \ds~satellites by their parent halo mass, showing any differences in \rHI~when controlled for \mHI.  The running medians for each halo mass bin give a hint of a trend that higher halo mass means slightly lower \rHI~for fixed $m_{\rm H\,{\LARGE{\textsc i}}}\!>\!10^{8.5}\,{\rm M}_{\odot}$.  Specifically, the separation between the lowest and highest halo mass bins reaches a maximum of $\sim$0.03\,dex.
Similar behaviour to a lesser extent is seen for the upper percentiles.  Only for the lower percentiles is there a more noticeable separation, but this becomes less clear for $M_{\rm halo} \! \gtrsim \! 10^{14}\,{\rm M}_{\odot}$; at these masses, the lower percentiles appear to be more convergent.  In contrast to the medians, the separation of these percentiles between the lowest and highest halo mass bins exceeds 0.08\,dex when $m_{\rm H\,{\LARGE{\textsc i}}}\!>\!10^9\,{\rm M}_{\odot}$.

So how does this low-\rHI~population fit in with the picture of Fig.~\ref{fig:trunc}?  The short answer: these galaxies tend to be those with higher central gas surface densities.  This is exemplified by the $\bar{\Sigma}_0\!=\!9$ (black) and $\bar{\Sigma}_{\rm 0,H}\!=\!150$ (cyan) curves in Fig.~\ref{fig:trunc}.  
To explain: this population starts slightly on the lower side of the size--mass relation, experiences a minimal horizontal evolution once truncation starts, and then begins to move further down and away from the primary relation.  Physically, the \HI~gets reduced to that in the densest allowable state.  Higher average density implies lower $r^2_{\rm H\,{\LARGE{\textsc i}}}/m_{\rm H\,{\LARGE{\textsc i}}}$.  The further along the truncation tracks in Fig.~\ref{fig:trunc} the galaxies move, the stronger the ram pressure they must be feeling, and therefore the more massive a halo they must reside in.  Based on our results, the corresponding halo masses required to move galaxies along the concave-up parts of those tracks (until their gradients reach their minimum) should continuously cover the range from $\lesssim\!10^{12}$ to $\lesssim\!10^{14}\,{\rm M}_{\odot}$.  In haloes of greater mass, stripping must be sufficiently strong to take galaxies beyond this, where the tracks have a constant or slow-changing gradient (for models 1/2 and 3, respectively).  From here, further truncation from more-massive haloes has zero or little effect on their displacement from the nominal \HI~size--mass relation, and thus the lower percentiles in the lower panel of Fig.~\ref{fig:SizeMassDS} become converged.

The overarching conclusion here is that galaxy environment indeed (only) plays a second-order role in the \HI~size--mass relation.  We examine this concept further, under a different definition of \HI~size, in Appendix \ref{sec:rhalf}.


\subsection{Results from TNG100}

The works of \citet{stevens19} and \citet{diemer19} have shown that the \HI~properties of galaxies in the TNG100 simulation at \zo~broadly align with observations.  This is true when galaxies are broken into centrals and satellites, and further when satellites are broken into bins of parent halo mass \citep{stevens19}.  This allows us to conclude that the effects of ram-pressure stripping in the simulation generally represent reality.  This is supported by the analysis of jellyfish galaxies in TNG by \citet{yun19}.  With this in mind, we can use TNG100 as a second, independent test of whether a galaxy's environment plays any role in where it sits in the \HI~size--mass plane.  What makes this test independent is that, because TNG100 is a hydrodynamic simulation, hydrodynamical and gravitational effects like ram-pressure and tidal stripping self-consistently result from interactions calculated at the simulation's smallest resolvable scale, meaning they do not need to be modelled explicitly.  The simulation is therefore agnostic \emph{a priori} (and predictive \emph{a posteriori}) as to how satellite stripping functions on a macroscopic scale, such as whether disc truncation is sufficiently descriptive or not.

\begin{figure}
\centering
\includegraphics[width=\textwidth]{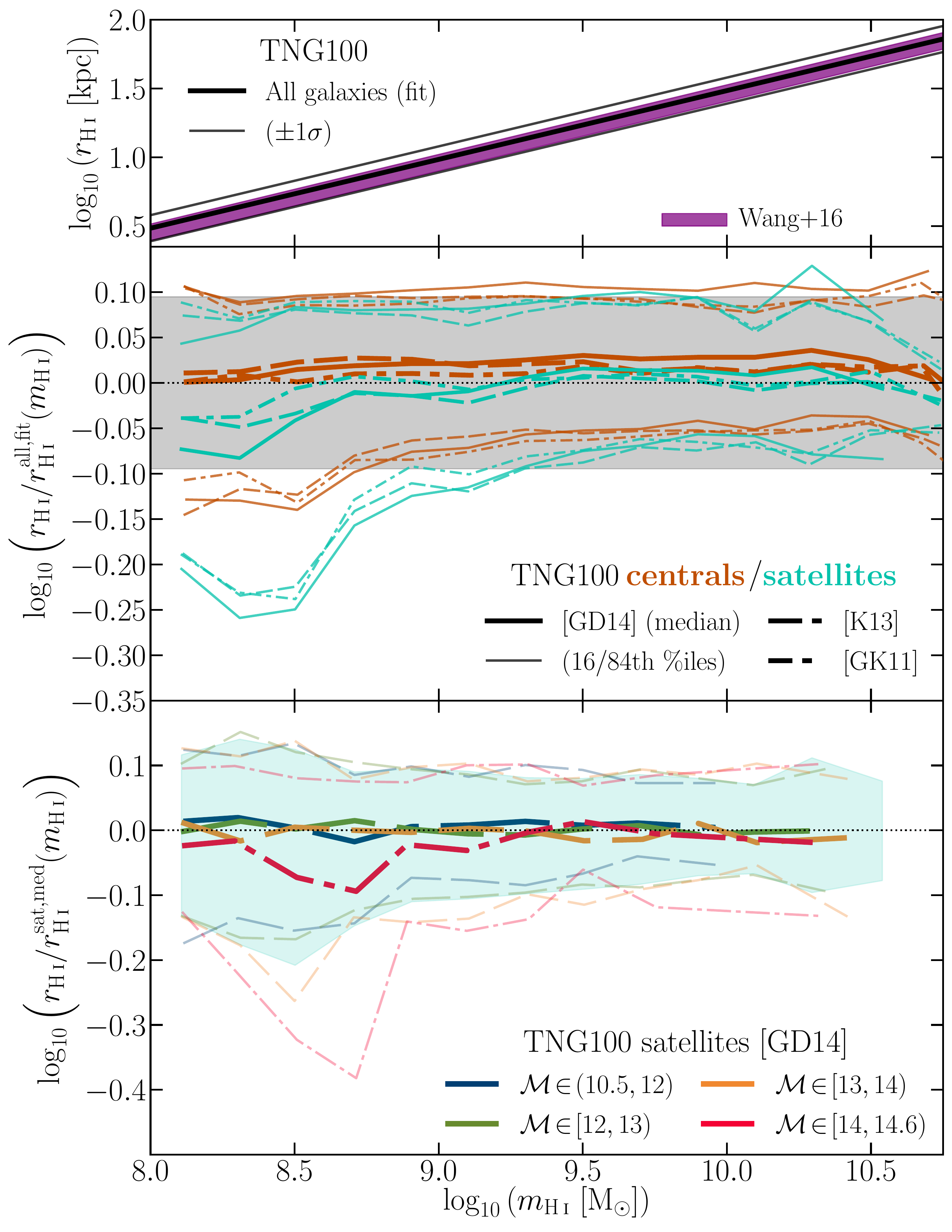}
\caption{As for Fig.~\ref{fig:SizeMassDS} but now assessing TNG100 galaxies at \zo.  Only galaxies with $m_*\!\geq\!10^9\,{\rm M}_{\odot}$ are included \citep[following the sample in][]{stevens19}.  Line styles in the second panel indicate the post-processing prescription used for the \HI/\Htwo~breakdown, which give effectively identical results.  Line styles in the bottom panel instead correspond to the range of satellites' host halo masses; for clarity, we only show the \citet{gd14} prescription here, as results from the other prescriptions are again very similar.}
\label{fig:SizeMass}
\end{figure}

With Fig.~\ref{fig:SizeMass}, we repeat the process done for \ds~in the previous subsection.  That is, we first plot the best-fitting fixed-slope \HI~size--mass relation for TNG100 galaxies in the top panel, then show potential deviations from this for satellites and centrals separately in the middle panel, and finally show the secondary effect of parent halo mass for satellites in the bottom panel (using the same halo mass bins as in \citealt{stevens19}).  In the middle panel, we show results for three prescriptions for separating the neutral gas in the simulation into its atomic and molecular components.  The results from all three are barely distinguishable, which is why we only show one prescription in the bottom panel (cf.~the results in \citealt{stevens19}).  In fact, centrals and satellites are barely distinguishable from each other either, in line with the results of \ds.

TNG100 has another feature similar to \ds~in that as one approaches low \HI~masses ($\sim\!10^8\,{\rm M}_{\odot}$), the simulated galaxies obey the \HI~size--mass relation less strictly, and centrals and satellites start behaving slightly differently.  Again, we heed caution in reading too much into this, as \rHI~is not always well resolved for these galaxies; while we have imposed a minimum \rHI~equal to the minimum gravitational softening scale for gas in the simulation, for \rHI~to be \emph{well} resolved would require it to be at least \emph{several} times the softening scale (i.e.~$>\!1\,{\rm kpc}$).

As for the finer impact of environment, the median \rHI~of TNG100 satellites at fixed $m_{\rm H\,{\LARGE{\textsc i}}} \! > \! 10^9\,{\rm M}_{\odot}$ is practically independent of halo mass.  That in itself is consistent with the earlier results in this section, but one notable difference in Fig.~\ref{fig:SizeMass} is a drop in $\Delta\log_{10}\!\left(r_{\rm H\,{\LARGE{\textsc i}}} \right)$ at $m_{\rm H\,{\LARGE{\textsc i}}} \! \simeq \! 10^{8.5}\,{\rm M}_{\odot}$ for satellites only in haloes of $M_{\rm 200c} \! \geq \! 10^{14}\,{\rm M}_{\odot}$.  This is seen most obviously in the median and 16th percentile lines.  Although potentially interesting, this should be taken with a grain of salt; in addition to the resolution limitations mentioned above, TNG100 only has 14 haloes at these masses (and it has zero with $M_{\rm 200c} \! > \! 10^{14.6}\,{\rm M}_{\odot}$, which is why there is one mass bin fewer in Fig.~\ref{fig:SizeMass} than Fig.~\ref{fig:SizeMassDS}).  There are also fewer total satellites (that contribute to Fig.~\ref{fig:SizeMass}) in this halo mass bin (299) than the others.
Otherwise, there is once again a divide in the lower percentiles for satellites in the lowest and highest halo mass bins, although this is less clean that it was for \ds.  We have confirmed that the TNG100 galaxies with lower \rHI~values are those with the highest $\bar{\Sigma}_0$ fits for models 1 and 2.  Again then, any effect environment has on the \HI~size--mass relation is secondary.


\section{Conclusion}
\label{sec:conc}

That the \HI~size--mass relation is so tight is perhaps unsurprising.  Given the commonality of how \HI~is distributed in most galaxies, and the tendency for \HI~to saturate due to the \HI--\Htwo~phase transition, it is a natural consequence that $r_{\rm H\,{\LARGE{\textsc i}}} \! \approxprop \! m_{\rm H\,{\LARGE{\textsc i}}}^{0.5}$  with a small scatter (Section \ref{sec:models}; also see \citealt{wang14,wang16}).

We have demonstrated analytically and with two different cosmological-simulation methods that satellite galaxies are no different to centrals in their \HI~size--mass relation to first order (Section \ref{sec:env}).  Effects such as ram-pressure stripping cause galaxies to move predominantly down and along the relation; which specific galaxies lie above or below the median (or best-fitting) relation might change, but the scatter and median remain effectively unchanged, with only the lower tail of the size distribution at fixed mass dragged down by $\lesssim\!0.1$\,dex.

The conclusions of this paper are applicable to galaxies with $m_* \! \geq \! 10^9\,{\rm M}_{\odot}$ and $m_{\rm H\,{\LARGE{\textsc i}}} \! \geq \! 10^8\,{\rm M}_{\odot}$, per our simulation mass limitations.  Given the mass range of the observations we have assessed ($m_{\rm H\,{\LARGE{\textsc i}}} \! \gtrsim \! 10^{6.3}\,{\rm M}_{\odot}$), these feasibly could extend to lower masses too.  
We have demonstrated that selecting galaxies in a fixed bracket of \HI~richness, morphology, or star formation activity does not change the crux of our results, even if the \emph{exact} parameters (most notably the scatter) of the best-fitting \HI~size--mass relation to a sample of galaxies is susceptible to biases in these properties (Section \ref{ssec:type}).
While we have focussed on galaxies at \zo, our conclusions should be qualitatively applicable across a wide redshift range (although there may be small systematics related to redshift that we have not explored -- see e.g.~fig.~7 of \citealt{ob09}).  
This gives promise that an \HI~size can be accurately inferred from single-dish or unresolved 21-cm detections.  This is important for large \HI~surveys like WALLABY\footnote{Wide-field {\sc Askap} L-band Legacy All-sky Blind surveY (Koribalski et al.~in preparation)} and APERTIF,\footnote{APERture Tile In Focus} as most detected galaxies will not have directly resolved \HI~sizes.

The robustness of the \HI~size--mass relation makes it an obvious test for any model or simulation of galaxy evolution.  It should be difficult to get the slope wrong by more than a few per cent, the scatter by more than a factor of $\sim$2, and the normalization wrong by more than the scatter's magnitude.
Any large tension with the observed \HI~size--mass relation should therefore provide motivation to revise feedback models and/or assumptions about the interstellar medium.
In practice, we found no impact from the way the \HI-to-\Htwo~ratio is treated in TNG100 (cf.~Fig.~\ref{fig:SizeMass} of this paper and fig.~5 of \citealt{diemer19}).  A similar conclusion can be drawn for \ds~(cf.~Fig.~\ref{fig:SizeMassDS} of this paper and fig.~3 of \citealt{lutz18}).

Even if feedback (or any process) were to generate a `hole' in the centre of an \HI~disc, unless that hole were sufficiently large to qualify the galaxy as a ring galaxy (and, perhaps, even then), it would \emph{still} lie on the observed \HI~size--mass relation.  This simply arises from the multiplicative $r$ term in the integrand used for calculating a galaxy's \HI~mass, meaning the central region only contributes a small percentage to the integral.

There is nothing mystical about the \HI~size--mass relation.  It is inevitable.


\section*{Acknowledgements}
All plots in this paper were built with the {\sc matplotlib} package for {\sc python} \citep{hunter07}.   
ARHS thanks G.~Kauffmann for discussion and funding to visit MPA that helped facilitate some of this work, C.~Howlett for practical help with some of the mathematics in this paper, and the IllustrisTNG team for access to the simulation data.
Parts of this research were supported by the Australian Research Council Centre of Excellence for All Sky Astrophysics in 3 Dimensions (ASTRO 3D), through project number CE170100013 .
FM acknowledges support through the Program `Rita Levi Montalcini' of the Italian MIUR.


\appendix

\section{Model parameter distribution functions}
\label{app:dists}

As discussed in Section \ref{ssec:infer}, a key part in understanding the precise normalization and scatter of the \HI~size--mass relation lies in the probability distributions of parameters that describe galaxies' \HI~surface density profiles.  While we could not directly predict these from analytic modelling, we were able to obtain said distributions as an outcome from fitting our analytic model $\bar{\Sigma}_{\rm H\,{\LARGE{\textsc i}}}(\bar{r})\,\bar{r}$ profiles to observations (important percentiles were given in Table \ref{tab:params}).  This same exercise can be done for \ds~and TNG100 galaxies too.

\begin{figure}
\centering
\includegraphics[width=\textwidth]{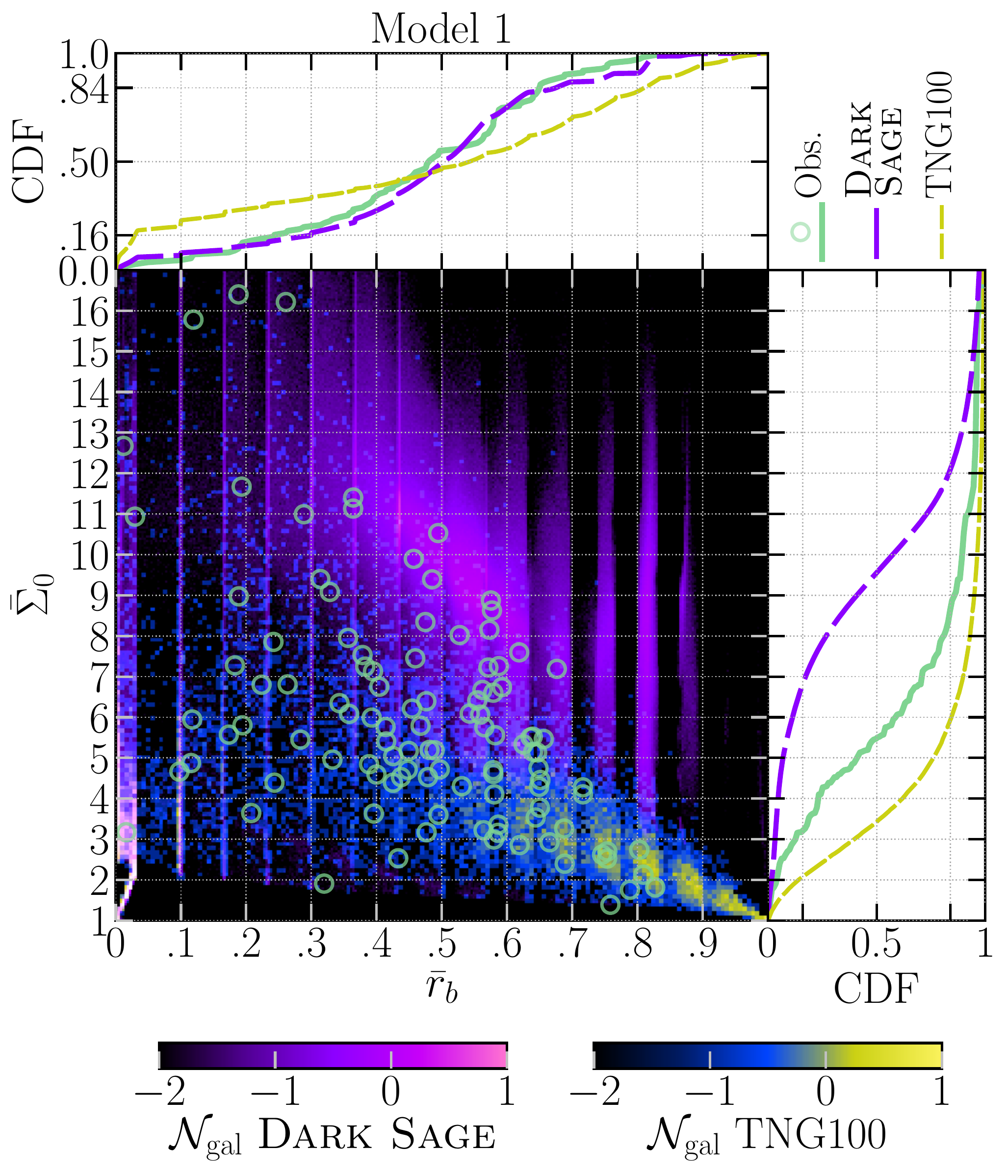}
\caption{Main panel: Two-dimensional distribution functions of model 1's parameter space from the \HI~surface density profile fits to observations and simulations (see Section \ref{sec:models}).  Pixels are coloured according the number density of \ds~and/or TNG100 galaxies; $\mathcal{N}_{\rm gal}$ is the base-10 logarithm of the fraction of total galaxies per unit square on the axes.  Smaller pixels are used for \ds, as there are many more galaxies than TNG100.  Where galaxies from both simulations occupy the same area of parameter space, the RGB colours from the two colour bars are summed.  The corrugated structure is an artefact of our \SHIr~profiles all using a common grid of fixed bin width in $\bar{r}$.  Circles are individual observations, the same as in Fig.~\ref{fig:scattermap}.  Smaller panels: Cumulative distribution functions for each model parameter from the same data.  Key percentiles for observations are summarized in Table \ref{tab:params}.}
\label{fig:model1dist}
\end{figure}

\begin{figure}
\centering
\includegraphics[width=\textwidth]{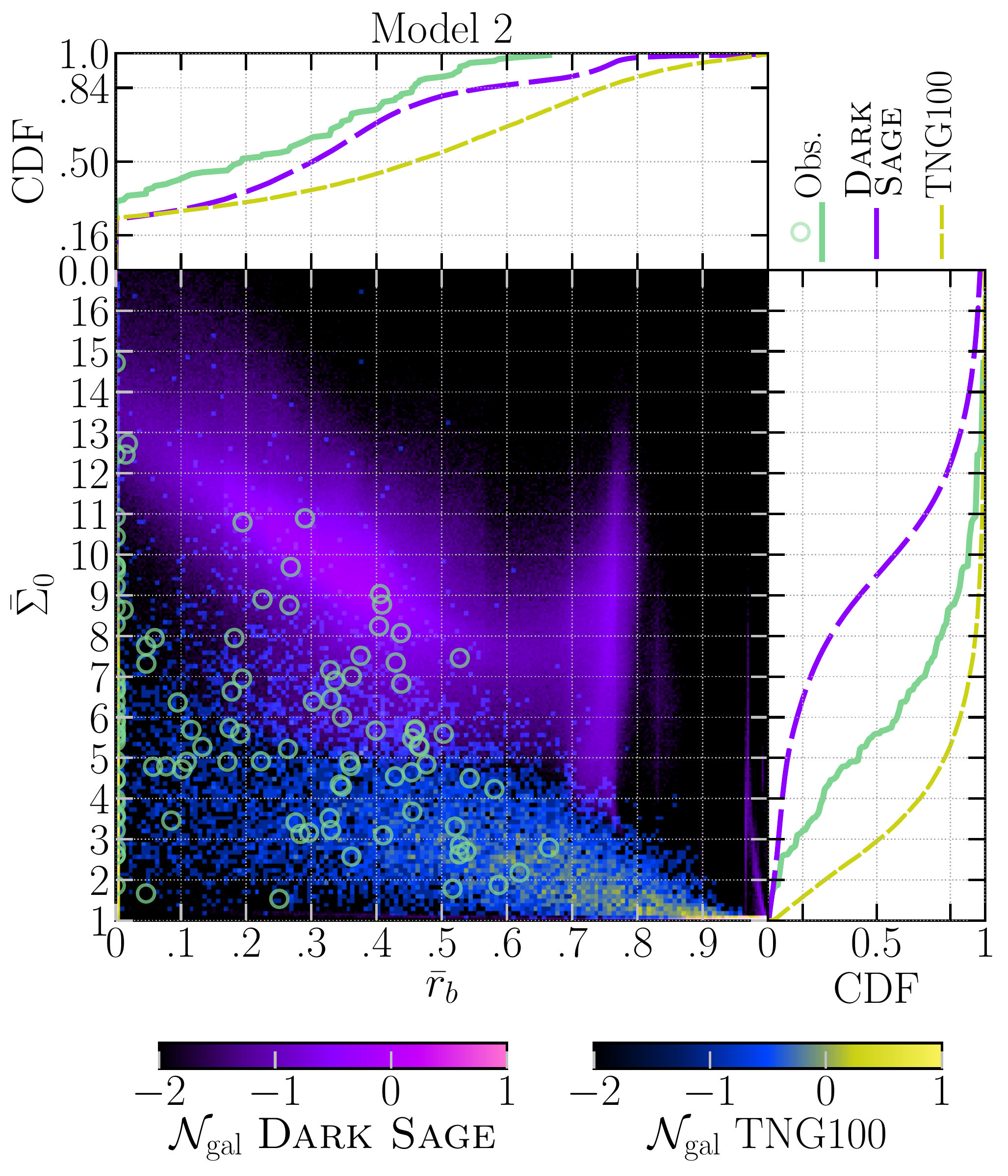}
\caption{As per Fig.~\ref{fig:model1dist} but now for model 2's parameter space.}
\label{fig:model2dist}
\end{figure}

\begin{figure}
\centering
\includegraphics[width=\textwidth]{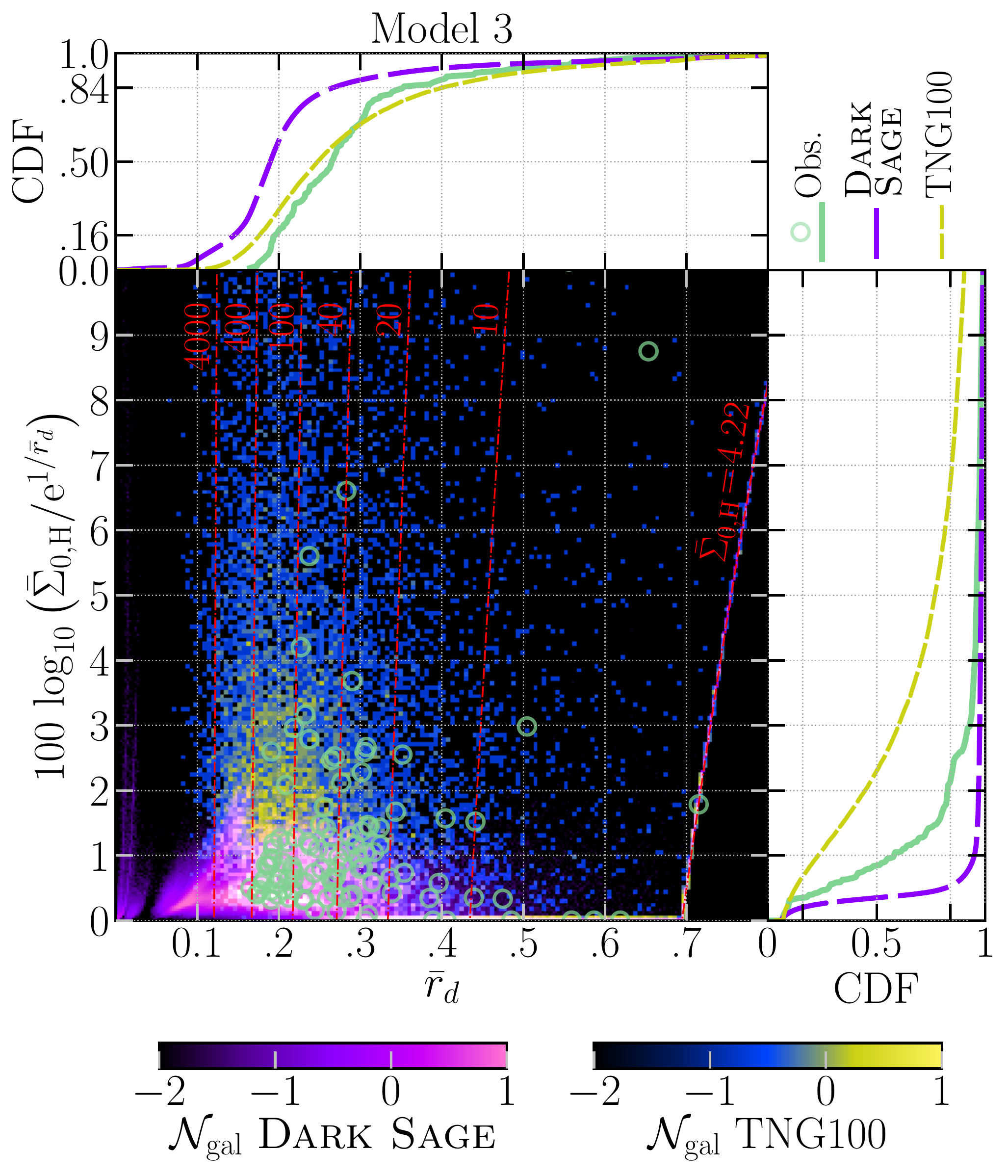}
\caption{As per Fig.~\ref{fig:model1dist} but now for model 3's parameter space.  The thin, dot-dashed lines give example reference values of $\bar{\Sigma}_{\rm 0,H}$.}
\label{fig:model3dist}
\end{figure}

In Figs \ref{fig:model1dist}--\ref{fig:model3dist}, we present the two-dimensional probability distribution functions of our three model parameter spaces, based on the fits to each of our three data sources.  These figures also include the cumulative distribution functions of each individual parameter.
There are varying degrees of similarity and difference in the observation and simulation parameter distributions.  For example, \ds~has systematically higher $\bar{\Sigma}_0$ (for both models 1 and 2) for its galaxies versus observations, which is in line with \ds~discs generally having too many baryons in their centres \citep[for discussion on this, see][]{stevens16,stevens18}.  Given that the \HI~size--mass ratio is less sensitive to variations in $\bar{\Sigma}_0$ for higher initial values of $\bar{\Sigma}_0$ (Fig.~\ref{fig:deriv}), it makes sense that \ds~has a smaller scatter in the \HI~size--mass relation than what is observed (Table \ref{tab:relations}).  Likewise, because TNG100 galaxies tend to have low $\bar{\Sigma}_0$, it follows that the simulation has a larger scatter in the \HI~size--mass relation.  Similarly, the limited (extended) range of model-3 $\bar{\Sigma}_{\rm 0,H}$ fits for \ds~(TNG100) also implies a smaller (larger) scatter in the \HI~size--mass relation relative to observations.  Broadly speaking, there is less variation in the distributions of $\bar{r}_b$ and $\bar{r}_d$ between the datasets.

Where Fig.~\ref{fig:model1dist} stands out from the others is in the fact that the distribution of model-1 parameter fits from the simulations has a corrugated structure.  In other words, both \ds~and TNG100 show common preferred values of $\bar{r}_b$ at regular intervals.  This is merely a reflection of the underlying radial grid used to build the \SHIr~profiles of the simulated galaxies (see Section \ref{ssec:sims}); $\bar{r}_b$ tends to be close to one of the points on that grid.  Evidently, there is a preferable number of points on any given \SHIr~profile that are deemed to be saturated, where small variations in $\bar{r}_b$ between the outermost saturated point and the next give only subtle changes to the goodness of fit to the rest of the profile.  Model 2 differs in that these small variations in $\bar{r}_b$ have a more significant effect on the rest of the profile, such that a better fit might be found with a different $(\bar{r}_b, \bar{\Sigma}_0)$ pair altogether.  There is no analogy to be drawn here for model 3, as its parameters are substantially different to models 1 and 2.

We emphasise that there are clear systematic differences in the parameter space occupancies of our two simulations.  Outside of similarities in how neutral gas is broken into atomic and molecular components, the way the interstellar media of galaxies is built and modelled in the two simulations is vastly different.  For TNG100, the gas structure of galaxies (i) is fully three-dimensional, (ii) self-consistently interacts with dark matter, (iii) is self-consistently affected by feedback, and (iv) has a quantitative consideration of temperature (which feeds into the phase decomposition).  Instead, \ds~gas discs (i) are modelled in one dimension, (ii) are built after the dark matter is evolved, (iii) only allow feedback to affect the same disc annulus where stars formed, and (iv) only treat the coldness of gas in a qualitative sense.  Bearing all of this in mind, it perhaps should not come as a surprise that there are systematic differences in the typical one-dimensional \HI~structure of galaxies predicted by these methods.


\section{Truncation equations}
\label{app:trunceq}

Here we provide the equations relating the \HI~size and mass of galaxies whose profiles have been truncated (see Section \ref{ssec:truncation}).
For model 1, after integrating \SHIr\,$r$~out to $r_t$ (using Equation \ref{eq:SHIr}), one obtains
\begin{multline}
m_{\rm H\,{\LARGE{\textsc i}}} = 2\pi\, \Sigma_0\, r_{\rm H\,{\LARGE{\textsc i}},init}^2 \bigg[ \frac{\bar{r}_b^2}{2} + \bar{r}_s(\bar{r}_s + \bar{r}_b) \\
 - \bar{r}_s(\bar{r}_s + \bar{r}_t)\exp\left(\frac{\bar{r}_b-\bar{r}_t}{\bar{r}_s}\right) \bigg]\,.
\label{eq:mtrunc1}
\end{multline}
This equation holds for $\bar{r}_t \! > \! \bar{r}_b$.  Physically, a case where $\bar{r}_t \! \leq \! \bar{r}_b$ is the same type of profile as one with $\bar{r}_b \! = \! 1$; i.e.~\SHIr~is constant until a radius where it drops to zero, meaning $m_{\rm H\,{\LARGE{\textsc i}}} \! = \! \pi \Sigma_0 r_{\rm H\,{\LARGE{\textsc i}}}^2 \!= \! \pi \Sigma_0 r_{\rm trunc}^2$.  

An equivalent form of Equation (\ref{eq:mtrunc1}) is also easily found for model 2:
\begin{multline}
m_{\rm H\,{\LARGE{\textsc i}}} = \pi\, \Sigma_0\, r_{\rm H\,{\LARGE{\textsc i}},init}^2 \bigg[ \bar{r}_b^2 + \sqrt{\pi}\, \bar{r}_S\, \bar{r}_b\, {\rm erf}\left(\frac{\bar{r}_t-\bar{r}_b}{\bar{r}_S}\right) \\
+ \bar{r}_S^2 - \bar{r}_S^2 \exp\left(-\frac{(\bar{r}_b-\bar{r}_t)^2}{\bar{r}_S^2} \right) \bigg]\,,
\label{eq:mtrunc2}
\end{multline}
where `erf' is the Gauss error function.  

The same procedure for model 3 gives
\begin{subequations}
\label{eq:mtrunc3}
\begin{multline}
m_{\rm H\,{\LARGE{\textsc i}}} = 2\pi\, \Sigma_{\rm 0,H}\, \bar{r}_d\, r_{\rm H\,{\LARGE{\textsc i}},init}^2\, \bigg[\frac{5}{3}\, B_1\, H_1\, \bar{r}_t \\
 + \frac{25}{9}\, \left(B_2\,H_2 - B_1\,H_3 \right)\, \bar{r}_d \bigg]\, \bigg[-B_2\, B_0\bigg]^{-1}\,,
\end{multline}
\begin{equation}
B_0 \equiv \exp\left(\frac{1.6}{\bar{r}_d}\right) - \Sigma_{\rm 0,H}\, \exp\left(\frac{0.6}{\bar{r}_d}\right)\,,
\end{equation}
\begin{multline}
B_1 \equiv \exp\left( \frac{2.2\,\bar{r}_t}{\bar{r}_d} \right) - \exp\left( \frac{1.6 + 0.6\,\bar{r}_t}{\bar{r}_d} \right) \\
 +  \Sigma_{\rm 0,H}\, \exp\left(\frac{0.6+0.6\,\bar{r}_t}{\bar{r}_d}\right)\,,
\end{multline}
\begin{equation}
B_2 \equiv B_3 - B_0\,,
\end{equation}
\begin{equation}
B_3 \equiv \exp\left(\frac{1.6\,\bar{r}_t}{\bar{r}_d} \right)\,,
\end{equation}
\begin{equation}
H_1 \equiv\, _2F_1 \left(0.375,\, 1;~1.375;~B_3\,B_0^{-1} \right)\,,
\end{equation}
\begin{equation}
H_2 \equiv\, _3F_2 \left(1,\, 0.375,\, 0.375;~1.375,\,1.375;~B_0^{-1} \right)\,,
\end{equation}
\begin{equation}
H_3 \equiv\, _3F_2 \left(1,\, 0.375,\, 0.375;~1.375,\,1.375;~B_3\,B_0^{-1} \right)\,,
\end{equation}
\end{subequations}
where the~$_pF_q\left(\alpha_1,...,\alpha_p;~\beta_1,...,\beta_q;~\gamma \right)$ terms are hypergeometric functions (note that these are \emph{not} regularized like Equation \ref{eq:m3sm}).


\section{An alternative \HI~size measure}
\label{sec:rhalf}

\begin{figure*}
\centering
\includegraphics[width=0.85\textwidth]{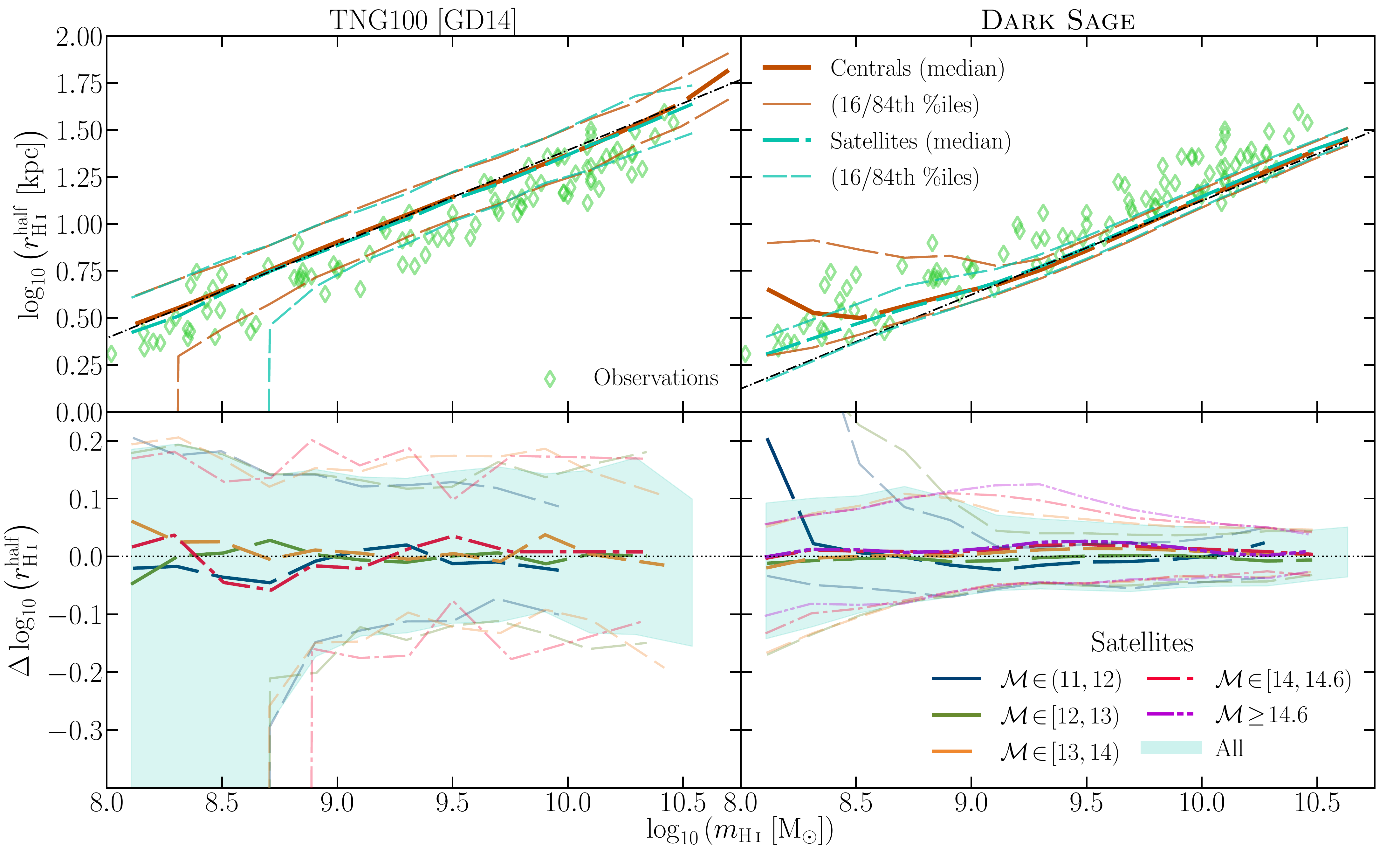}
\caption{Similar to Figs \ref{fig:SizeMassDS} \& \ref{fig:SizeMass} (for the right- and left-hand panels, respectively) but now the classical \HI~radius has been replaced with the \HI~half-mass radius.  For simplicity, we only show the \citetalias{gd14} prescription for the \HI/\Htwo~breakdown for TNG100 (the other prescriptions give effectively the same results).  Measurements of individual galaxies from our observed sample are overlaid in the top panels.  Thin, dot-dashed lines in the top panel give the best-fitting linear relation to the simulated galaxies, assuming a slope of 0.5.  Systematic differences seen between the simulations and observations are consistent with the standard \HI~size--mass relation results.}
\label{fig:rhalf}
\end{figure*}

It is a fair question to ask whether part of the tightness and ubiquity of the \HI~size--mass relation comes from how \rHI~is defined.  As mentioned in Section \ref{sec:intro}, the definition $\Sigma_{\rm H\,{\LARGE{\textsc i}}}(r_{\rm H\,{\LARGE{\textsc i}}}) \! \equiv \! \Sigma_c \! = \! 1\,{\rm M}_{\odot}\,{\rm pc}^{-2}$ originates from the typical \HI~column density that radio observations have been sensitive to in the past (e.g.~\citealt{warmels88,broeils94}, where earlier works had to use $\bar{\Sigma}_c \! > \! 1$, e.g.~\citealt{bosma81b}).  While \citetalias{wang16} (see their fig.~3) have shown that varying $\Sigma_c$ by a factor of $\sim$3 has little impact on the interpretation of the \HI~size--mass relation (cf.~\citealt{begum08}, who use $\bar{\Sigma}_c \! < \! 1$), there is no fundamental physical reason why an absolute threshold surface density is the `right' way to measure \HI~size in the first place.  In the optical community, for example, a more common practice is to refer to the stellar size of a galaxy by its half-mass radius.  If, instead, we were also to use the radius enclosing half a galaxy's \HI~mass to define its \HI~size, would that significantly affect the \HI~size--mass relation?

We formally define \rhalf~through the expression
\begin{equation}
\frac{m_{\rm H\,{\LARGE{\textsc i}}}}{2}  = 2 \pi \int_0^{r_{\rm H\,{\LARGE{\textsc i}}}^{\rm half}} \Sigma_{\rm H\,{\LARGE{\textsc i}}}(r)\, r\, {\rm d}r\,.
\end{equation}
If we use model 1 or 2 to solve this integral, because \SHIr~is piecewise, we would need to separately consider the instances when $r_{\rm H\,{\LARGE{\textsc i}}}^{\rm half} \! \leq \! r_b$ and $r_{\rm H\,{\LARGE{\textsc i}}}^{\rm half} \! > \! r_b$; in the former case, one trivially obtains $m_{\rm H\,{\LARGE{\textsc i}}} \! = \! 2\pi\, \Sigma_0 [r_{\rm H\,{\LARGE{\textsc i}}}^{\rm half}]^2$. Otherwise, the solution to this integral for any of our models is essentially already given by Equations (\ref{eq:mtrunc1}--\ref{eq:mtrunc3}), where $r_t$ can be replaced with \rhalf.

The only real difference between the normal \HI~size--mass relation and the \mHI--\rhalf~relation is that the latter is more sensitive to the \HI~profile parameter values.  Predictably, it should then have a larger scatter (but not too much larger), a lower normalization, and the same slope.

To test these expectations, we measure \rhalf~from observed, TNG100, and \ds~galaxies, plotting them against \mHI~in Fig.~\ref{fig:rhalf}.
One clear difference for the \ds~galaxies in Fig.~\ref{fig:rhalf} versus Fig.~\ref{fig:SizeMassDS} is the upturn in the typical \rhalf~values of centrals at $m_{\rm H\,{\LARGE{\textsc i}}} \! \lesssim \! 10^9\,{\rm M}_{\odot}$.  This is not a sign that the analytic model breaks down; rather, what we are seeing here is a tendency for low-\mHI~galaxies to have both low $\bar{r}_b$ and low $\Sigma_0$.  This could potentially just be a reflection of the fact that this mass scale is at the simulation's resolution limit; the median \HI~mass of \ds~galaxies occupying Millennium haloes of 100 particles is $\sim\!10^9\,{\rm M}_{\odot}$.  Our \ds~galaxy sample with $m_{\rm H\,{\LARGE{\textsc i}}} \! \lesssim \! 10^9\,{\rm M}_{\odot}$ is therefore almost certainly not halo-mass complete.  Current results from these galaxies should thus be taken with a grain of salt, but they should mature as \ds~is transitioned to higher-resolution simulations. TNG100 does \emph{not} share the same feature.  That is, satellites and centrals follow the same power-law-like relation for the full \HI~mass range (similar to the top panel of Fig.~\ref{fig:SizeMass}).

For \ds, satellites in low-mass haloes behave the same as centrals in their upturn in \rhalf~at low \mHI~(cf.~the top and bottom panels of Fig.~\ref{fig:rhalf}).  But for most satellites, i.e.~those in haloes of $M_{\rm 200c}\!\gtrsim\!10^{13}\,{\rm M}_{\odot}$, there is no strong upturn.  To explain this, we need to understand how disc truncation would affect the \mHI--\rhalf relation.  As pointed out in Section \ref{ssec:truncation}, a truncated model-1 or -2 profile with $\bar{r}_t \! < \! \bar{r}_b$ is indistinguishable from a non-truncated profile with $\bar{r}_b \! = \! 1$.  The higher the halo mass, the more truncated the satellite's \HI~is, therefore the larger the typical fitted $\bar{r}_b$ is, meaning the more common it is for $r_{\rm H\,{\LARGE{\textsc i}}}^{\rm half} \leq r_b$, where $r_{\rm H\,{\LARGE{\textsc i}}}^{\rm half} \propto m_{\rm H\,{\LARGE{\textsc i}}}^{0.5}$.  Also, because many of these satellites will have had more \HI~before infall, they are less likely to be biased towards low $\Sigma_0$ like \ds~centrals of the same current mass; in principle, unlike $\bar{r}_b$, truncation should not affect the best-fitting $\Sigma_0$ to \SHIr~[or, in this case, $\bar{\Sigma}_{\rm H\,{\LARGE{\textsc i}}}(\bar{r})\,\bar{r}$].  Both effects mean satellites at low \mHI~should typically have lower \rhalf~than centrals.

The effect of halo mass on \rhalf~is otherwise the \emph{opposite} to how it was for \rHI.  That is, at fixed $m_{\rm H\,{\LARGE{\textsc i}}} \! \in \! [10^9,10^{10}]\,{\rm M}_{\odot}$, satellites in higher halo masses have slightly higher \rhalf~on average.  This can again be explained in terms of $\Sigma_0$. \ds~satellites in higher-mass haloes tend to have slightly lower $\Sigma_0$; the medians for those in haloes of $M_{\rm 200c} \! < \! 10^{12}\,{\rm M}_{\odot}$ and $> \! 10^{14}\,{\rm M}_{\odot}$ are 10 and $9\,{\rm M}_{\odot}\,{\rm pc}^{-2}$, respectively.  While there is still a tendency for lower $\Sigma_0$ to also mean higher $\bar{r}_b$, the specific value of $\bar{r}_b$ is less important, as the vast majority of satellites have $r_{\rm H\,{\LARGE{\textsc i}}}^{\rm half} \! < \! r_b$ anyway.  What matters is $\Sigma_0 \! = \! \left\langle \Sigma_{\rm H\,{\LARGE{\textsc i}}}(<\!r_{\rm H\,{\LARGE{\textsc i}}}^{\rm half}) \right\rangle$, and a higher average density guarantees lower \rhalf~at fixed \mHI~by definition.

One reason \emph{why} \ds~satellites have lower $\Sigma_0$ in higher halo masses could be to do with coherent accretion.  Any gas that satellite galaxies accrete in \ds~is assumed to carry a constant specific-angular-momentum vector, fixed at infall.  Incoherent accretion leads to the build-up of more gas in the galaxy's centre; ergo, coherent accretion promotes lower $\Sigma_0$.  Satellites in more-massive haloes are likely to have been satellites for longer, and are therefore likely to have had more of their gas accreted coherently.  Because this is a feature of the model put in by hand, it is not obvious the extent to which the subtle impact environment has in the bottom panel of Fig.~\ref{fig:rhalf} should be reflected in reality.

The other opposite between the bottom panel of Fig.~\ref{fig:SizeMassDS} and bottom-right panel of Fig.~\ref{fig:rhalf} is that the effect of environment is more strongly seen in the lower percentiles of the former but the upper percentiles of the latter.  While the movement of galaxies in the classical \HI~size--mass plane away from the main relation was most significant for those with high $\Sigma_0$, it is the low-$\Sigma_0$ galaxies that are most sensitive movers in the \mHI--\rhalf~plane.

For TNG100, we note that any potential variation in satellites' \rhalf~with halo mass is even less evident than that seen for \rHI~in the bottom panel Fig.~\ref{fig:SizeMass}.  Consistent with expectation, the scatter in the \mHI--\rhalf~relation for TNG100 is larger than for \rHI, with a typical half-range between the 16th \& 84th percentiles of 0.13\,dex.  For \ds~galaxies with $m_{\rm H\,{\LARGE{\textsc i}}} \! > \! 10^9\,{\rm M}_{\odot}$, the same half-range is $\gtrsim\!0.05\,{\rm dex}$.

This exercise highlights that the universality and tightness of the \HI~size--mass relation is relatively insensitive to the definition of \HI~size.  It is a truly physical relation.

\label{lastpage}

\end{document}